\documentclass[%
 reprint,
 superscriptaddress,
 amsmath,amssymb,
 aps,
]{revtex4-1}

\usepackage {atbegshi}
\usepackage [dvipdfmx]{graphicx}
\usepackage{dcolumn}
\usepackage{bm}
\usepackage{hyperref}
\usepackage{txfonts}
\usepackage{braket}
\usepackage{multirow}
\usepackage {color}
\usepackage {empheq}

\begin{document}
\newcommand\red[1]{\textcolor{red}{#1}}
\newcommand\blu[1]{\textcolor{blue}{#1}}
\newcommand\mage[1]{\textcolor{magenta}{#1}}

\preprint{APS/123-QED}

\title{Chromatin state switching in a polymer model with mark-conformation coupling}


\author{Kyosuke Adachi}%
\email{kyosuke.adachi@riken.jp}
\affiliation{%
Nonequilibrium Physics of Living Matter RIKEN Hakubi Research Team, RIKEN Center for Biosystems Dynamics Research, 2-2-3 Minatojima-minamimachi, Chuo-ku, Kobe 650-0047, Japan
}%
\affiliation{%
RIKEN Interdisciplinary Theoretical and Mathematical Sciences Program, 2-1 Hirosawa, Wako 351-0198, Japan
}%
\author{Kyogo Kawaguchi}%
\email{kyogo.kawaguchi@riken.jp}
\affiliation{%
Nonequilibrium Physics of Living Matter RIKEN Hakubi Research Team, RIKEN Center for Biosystems Dynamics Research, 2-2-3 Minatojima-minamimachi, Chuo-ku, Kobe 650-0047, Japan
}%
\affiliation{%
RIKEN Cluster for Pioneering Research, 2-2-3 Minatojima-minamimachi, Chuo-ku, Kobe 650-0047, Japan
}%

\date{\today}

\begin{abstract}
We investigate the phase transition properties of the polymer-Potts model, a chain composed of monomers with magnetic degrees of freedom, with the motivation to study the conformation and mark switching dynamics of chromatin.
By the mean-field approximation, we find that the phase transition between the swollen-disordered state and the compact-ordered state is discrete; it is first-order as in the long-range Potts model, but with a significantly larger jump in magnetization (i.e., mark coherence) upon the ordering transition.
The results imply how small changes in epigenetic writer concentrations can lead to a macroscopic switching of the chromatin state, suggesting a simple mechanism of discrete switching observed, for instance, in cell differentiation.
\end{abstract}

\maketitle

\textit{Introduction.---}
Chromatin is a large polymer composed of monomers called nucleosomes, which are histone protein complexes wrapped with DNA~\cite{Cortini2016,Fierz2019}. 
The switching of cell states is encoded in the changes in epigenetics~\cite{Waddington1957} such as in the molecular and structrual changes in the chromatin.
The chromatin states have been traditionally categorized into two: euchromatin and heterochromatin, which correspond to active (open) and inactive (closed) parts of the chromatin in terms of gene expression and accessibility.
Consistent with this, recent chromosome conformation capture (Hi-C) experiments~\cite{Lieberman2009,Dixon2015} have identified the existence of two major compartments in the genome.
The regions within the same compartment share similar marks (i.e., chemical modifications) in the nucleosomes and tend to interact with each other more frequently than across~\cite{DiPierro2017}.
Experiments have shown that different marks on the nucleosome induce distinct interactions due to the natural attraction and repulsion between nuclesomes~\cite{Funke2016,Gibson2019}, or by mediating proteins such as HP1~\cite{Bannister2001,Machida2018}, and the polycomb repressive complexes~\cite{Angel2011,Boettiger2016}.

It has also been established that cell differentiation is accompanied by a large (megabase) scale  transition in the compartments as well as changes in the states of epigenetic marks~\cite{Dixon2015}.
The mechanism behind this switching, however, remains elusive.
Previous modeling studies have assumed mark-dependent interactions between nucleosomes in order to explain the observed contact maps and 3D structures of the chromatin~\cite{Jost2014NAR,Barbieri2012}. 
Other models have considered how the interaction between the marked histones lead to bistability in the coherent epigenetic marks~\cite{Dodd2007,Sneppen2008,Jost2014PRE}.
A natural question is then how the interplay of chromatin chain dynamics and the kinetics of nucleosome modifications can lead to the drastic switching of compartments observed in differentiation.

To model chromatin polymer dynamics under stochastic modifications of nucleosomes, the polymer-Potts model has been considered~\cite{Garel1999,Michieletto2016,Coli2018}.
In this model, the random motion of the polymer chain is accompanied by monomer-monomer interactions that depend on the histone marks, and the histone marks can stochastically switch due to enzymatic reactions and histone turnover~\cite{Deal2010, Dion2007}.
It has been numerically shown~\cite{Michieletto2016, Michieletto2018} that even for the Ising-type model, where there is essentially only two distinct states of the histone marks, there is a first-order-like transition between the swollen-disordered state, which corresponds to a loose polymer with spatially random marks, and the compact-ordered state, where the conformation is globular and the marks are coherent.
This abrupt transition is likely due to the coupling between the conformation change and the epigenetic switchings, although a concrete theory is still lacking.

In this paper, we investigate the phase transition properties of the polymer-Potts model by considering a polymer chain in continuum space with stochastic histone mark exchange.
Employing the Flory-type mean-field approximation for the dynamics of the chain, we write the pseudo free energy of the generic polymer-Potts model as a function of the order parameters representing the magnetization (i.e., mark coherence) and the polymer conformation.
For the Ising-type interaction, the transition between the swollen-disordered state and a compact-ordered state is first-order, consistent with simulations and theories investigating mark dynamics on self-avoiding random walks~\cite{Garel1999, Coli2018}.
In the general case with multiple types of marks, we find that the jump in magnetization at the transition point is always larger in the polymer-Potts model compared with the Potts-model counterpart, and also obtain a criteria for the absense of a continuous transition.
We further study the switching transition upon stretching of the chain, which serves as a simple model of force-induced epigenetic modification.

\begin{figure}[t]
\includegraphics[scale=0.35]{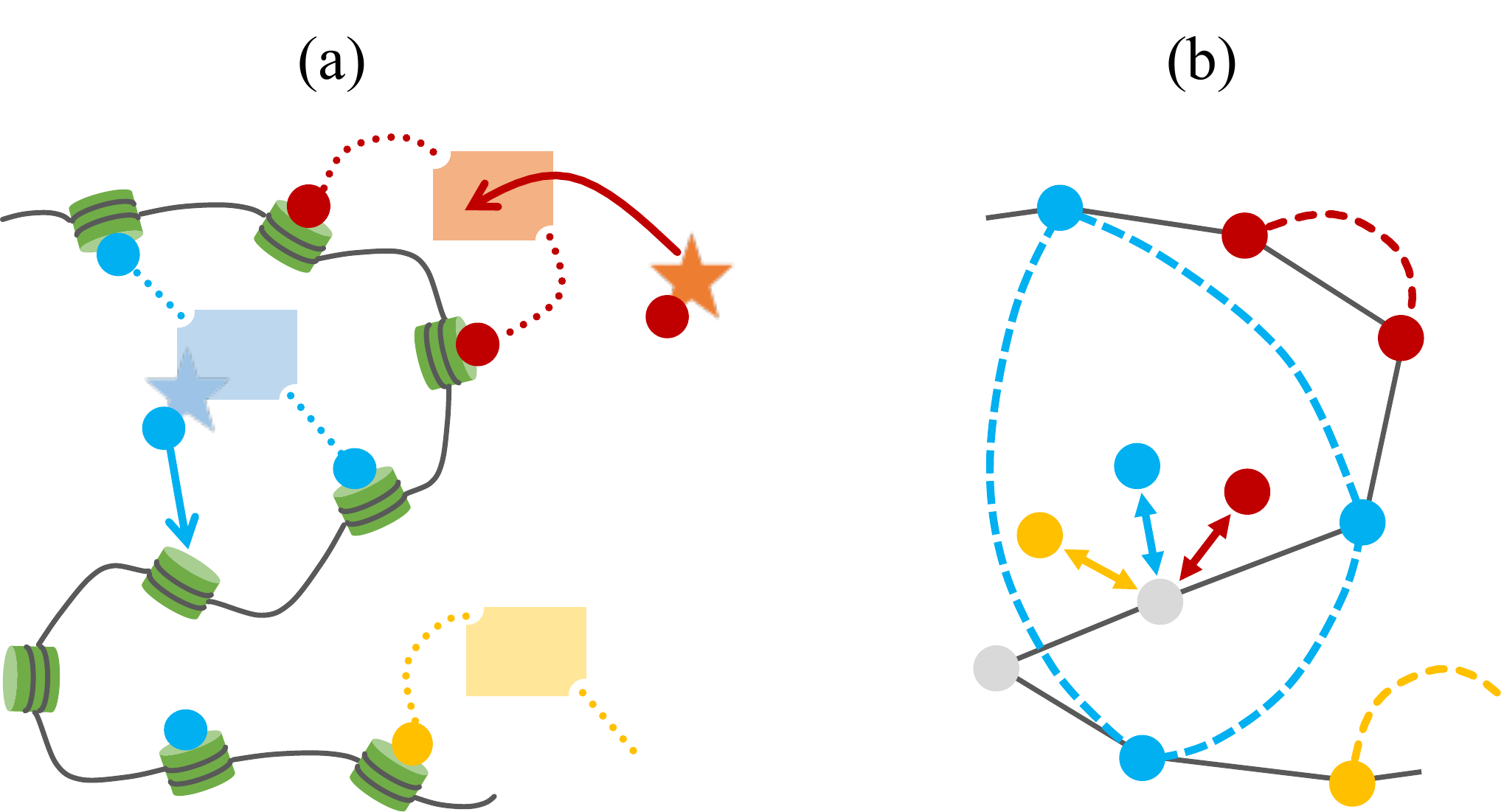}
\caption{(a)~Chromatin consists of a DNA (gray line) wrapping histones (green cylinders).
The readers (colored rectangles) connect histones with the same marks (colored circles), while the writers (colored stars) edit the histone marks.
(b)~The polymer-Potts model.
A monomer constituting the polymer has a changeable histone mark (colored circle), and there are mark-specific interactions between monomers (colored dashed lines, $J_{ij}$), in addition to the mark-independent interactions ($v$ and $w$).}
\label{Fig:Model}
\end{figure}

\textit{Model.---}
We consider a polymer model with Potts-like interactions between monomers.
The interactions are mediated in a histone-mark-dependent way by proteins that we call readers, and the marks can change stochastically due to enzymatic reactions caused by the writers (Fig.~\ref{Fig:Model}).
We assume that there are $q$ ($\geq 2$) types of marks.
Extending the Flory-type mean-field approximation~\cite{DeGennes1975} to the present situation, the pseudo free energy (per monomer) at a temperature $T$ reads
\begin{eqnarray}
f(\rho, \{ x_i \}) &=& v \rho + w \rho^2 - \rho \sum_{1\leq i,j \leq q} J_{ij} x_i x_j \nonumber\\
&& + \sum_{1 \leq i \leq q} \left( k_\mathrm{B} T x_i \ln x_i - h_i x_i \right).
\label{Eq:PseudoFreeEnergyDensity}
\end{eqnarray}
Here $\rho$ is the average monomer concentration given as $\rho \sim N / R^3$ with the polymer end-to-end distance $R$ and the total number of monomers $N$.
The other variables, $\{ x_i \}$, represent the $i$-th type mark occupation ($\sum_{i=1}^q x_i = 1$).
The first two terms in Eq.~\eqref{Eq:PseudoFreeEnergyDensity} correspond to the volume exclusion effect ($v, w > 0$).
Note that the parameters $v$ and $w$ represent the second and third virial coefficients, respectively.
The third term originates from the Potts-like two-body interactions between monomers mediated by the readers.
The detail of the interactions between the different types of marks is coded in $J_{ij}$, which is a real symmetric matrix.
The fourth term is the entropy associated with the mark degrees of freedom.
The last term represents the effect of external fields $\{ h_i \}$, which describes how much a specific epigenetic mark is favored, reflecting, for example, the concentration of the histone modification enzymes, i.e., the writers.
The equilibrium state is determined by minimizing Eq.~\eqref{Eq:PseudoFreeEnergyDensity} with respect to $\rho$ and $\{ x_i \}$.

A few remarks are to be made for Eq.~\eqref{Eq:PseudoFreeEnergyDensity}.
Firstly, fluctuation effects are neglected compared with the microscopic model, although its inclusion will likely not change the key results~\cite{Moore1977, Lifshitz1978}.
Secondly, while higher-order interaction terms are irrelevant near the conventional second-order coil-globule transition point~\cite{DeGennes1975}, inclusion of these terms may shift the transition point if the coil-globule transition becomes first-order, as in the situations explained below.
Nevertheless, we expect that Eq.~\eqref{Eq:PseudoFreeEnergyDensity} captures the key characters observed in simulations of similar systems~\cite{Michieletto2016, Coli2018} and is useful to generally analyse models with multiple kinds of marks.
Lastly, the chromatin state transitions of our interest is at the level of sub-regions of compartments or several topologically associated domains~\cite{Dixon2012}, which is megabase scale corresponding to $N=10^{3-5}$.
Although we have omitted all the terms that vanish in the limit of  $N \to \infty$ in Eq.~\eqref{Eq:PseudoFreeEnergyDensity}, it is straightforward to include higher-order terms and discuss their effects on the properties of transition~\cite{DeGennes1975,SM}.

\begin{figure}[t]
\includegraphics[scale=0.3]{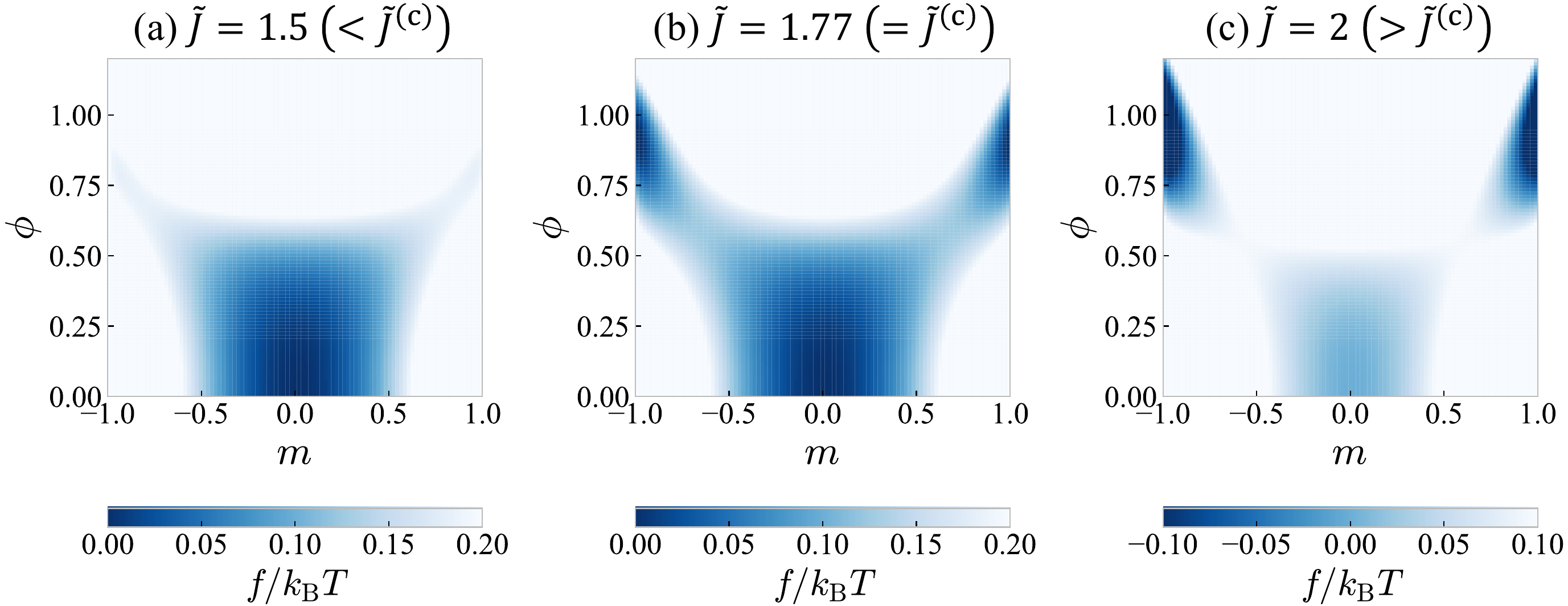}
\caption{Color plot of the pseudo free energy $f (\phi, m)$ [Eq.~\eqref{Eq:PseudoFreeEnergyDensity}] as a function of $\phi$, which represents the square root of the globule density, and the magnetization $m$ for the case of the Ising-type interaction with $\tilde{v} = 0.1$.
For (a)~$\tilde{J} < \tilde{J}^\mathrm{(c)}$, the minimum of $f$ is realized at $\phi = 0$ and $m = 0$ (swollen-disordered state), while for (c)~$\tilde{J} > \tilde{J}^\mathrm{(c)}$, it is realized at $\phi > 0$ and $|m| > 0$ (compact-ordered state).
There is bistability at (b)~$\tilde{J} = \tilde{J}^\mathrm{(c)}$, meaning that the transition between the swollen-disordered state and the compact-ordered state is discontinuous.}
\label{Fig:PseudoFreeEnergy}
\end{figure}

For the sake of understanding, let us first fix the mark degrees of freedom, $\{ x_i \}$.
Then, Eq.~\eqref{Eq:PseudoFreeEnergyDensity} is equivalent to the free energy of a classic polymer \cite{DeGennes1975} in the large $N$ limit.
As investigated in \cite{DeGennes1975}, there exists a transition between the coiled state, a swollen polymer with the average length scaling as $R \sim N^{3/5}$~\cite{Doi1986, DeGennes1979}, and the globule state, a densely packed polymer with $R \sim N^{1/3}$~\cite{DeGennes1975}, upon changing the overall two-body interaction (in the present case, $v - \sum_{i,j} J_{ij} x_i x_j$) from repulsive to attractive.
Such coil-globule transitions have been observed in experiments using DNA~\cite{Yoshikawa1996} and chromatin~\cite{Zinchenko2017}.
The coil-globule transition in this case is continuous in the limit of $N \to \infty$~\cite{DeGennes1975}, even beyond the mean-field approximation~\cite{Moore1977, Lifshitz1978}.

In another direction of simplification, we can consider the order-disorder transition of the marks under a fixed polymer density, $\rho$.
Assuming a globular configuration ($R \sim N^{1/3}$) and $J_{ij} = J \delta_{ij}$, Eq.~\eqref{Eq:PseudoFreeEnergyDensity} represents the mean-field free energy of the Potts model~\cite{Wu1982}.
In the limit of $N \to \infty$, the order-disorder transition is continuous for $q = 2$ while discontinuous for $q \geq 3$, which has been believed to be correct in three dimensions even beyond the mean-field approximation~\cite{Wu1982}.

\textit{Phase transition by interactions and fields.---}
Introducing the dimensionless globular order parameter $\phi := (w / k_\mathrm{B} T)^{1 / 4} \sqrt{\rho}$, 
we can express the equilibrium free energy as $\overline{f} = [\min_{\phi, \{ x_i \}} f(\phi, \{ x_i \}) \text{ s.t.} \sum_{i = 1}^q x_i = 1]$, where
\begin{eqnarray}
\frac{f(\phi, \{ x_i \})}{k_\mathrm{B} T} &=& \left (\tilde{v} -  \sum_{1\leq i,j \leq q} \tilde{J}_{ij} x_i x_j \right ) \phi^2 + \phi^4 \nonumber\\
&& + \sum_{1\leq i \leq q} \left( x_i \ln x_i - \frac{h_i x_i}{k_\mathrm{B} T} \right).
\label{Eq:PseudoFreeEnergyDensityPhi}
\end{eqnarray}
Here, the dimensionless two-body interaction strengths are defined as $\tilde{J}_{ij} := {J}_{ij}/ \sqrt{k_\mathrm{B} T w}$ and $\tilde{v} := v / \sqrt{k_\mathrm{B} T w}$.
As we have seen, if $\{ x_i \}$ or $\phi$ is fixed to some value, the system described by Eq.~\eqref{Eq:PseudoFreeEnergyDensityPhi} will show the conventional coil-globule or magnetic transition, respectively.

To see the effect of the coupling between $\{x_i\}$ and $\phi$, we first study the Ising-model case, $q = 2$ and $J_{ij} = J (2 \delta_{ij} - 1)$ with $h_i = 0$.
Introducing the magnetization, $m:= x_1 - x_2$, $f(\phi, m)$ for the case of $\tilde{v} = 0.1$ is shown in Fig.~\ref{Fig:PseudoFreeEnergy}.
Let us denote the equilibrium values of $\phi$ and $m$ as $\phi^*$ and $m^*$, respectively.
In this model, there is a critical value $\tilde{J}^\mathrm{(c)}$ such that for $\tilde{J} < \tilde{J}^\mathrm{(c)}$, the swollen-disordered phase is the equilibrium [$\phi^* = 0$ and $m^* = 0$, Fig.~\ref{Fig:PseudoFreeEnergy}(a)], whereas for $\tilde{J} > \tilde{J}^\mathrm{(c)}$, this switches to the compact-ordered state [$\phi^* > 0$ and $|m^*| > 0$, Fig.~\ref{Fig:PseudoFreeEnergy}(c)].
At the transition point [$\tilde{J} = \tilde{J}^\mathrm{(c)}$, see Fig.~\ref{Fig:PseudoFreeEnergy}(b)], both the swollen-disordered state and the compact-ordered state are stable, meaning that there is a first-order transition. 
Thus, a switching transition can occur by simply changing the strength of the reader-mediated interaction, as has been seen numerically in a similar model~\cite{Michieletto2016}.

\begin{figure}[t]
\includegraphics[scale=0.29]{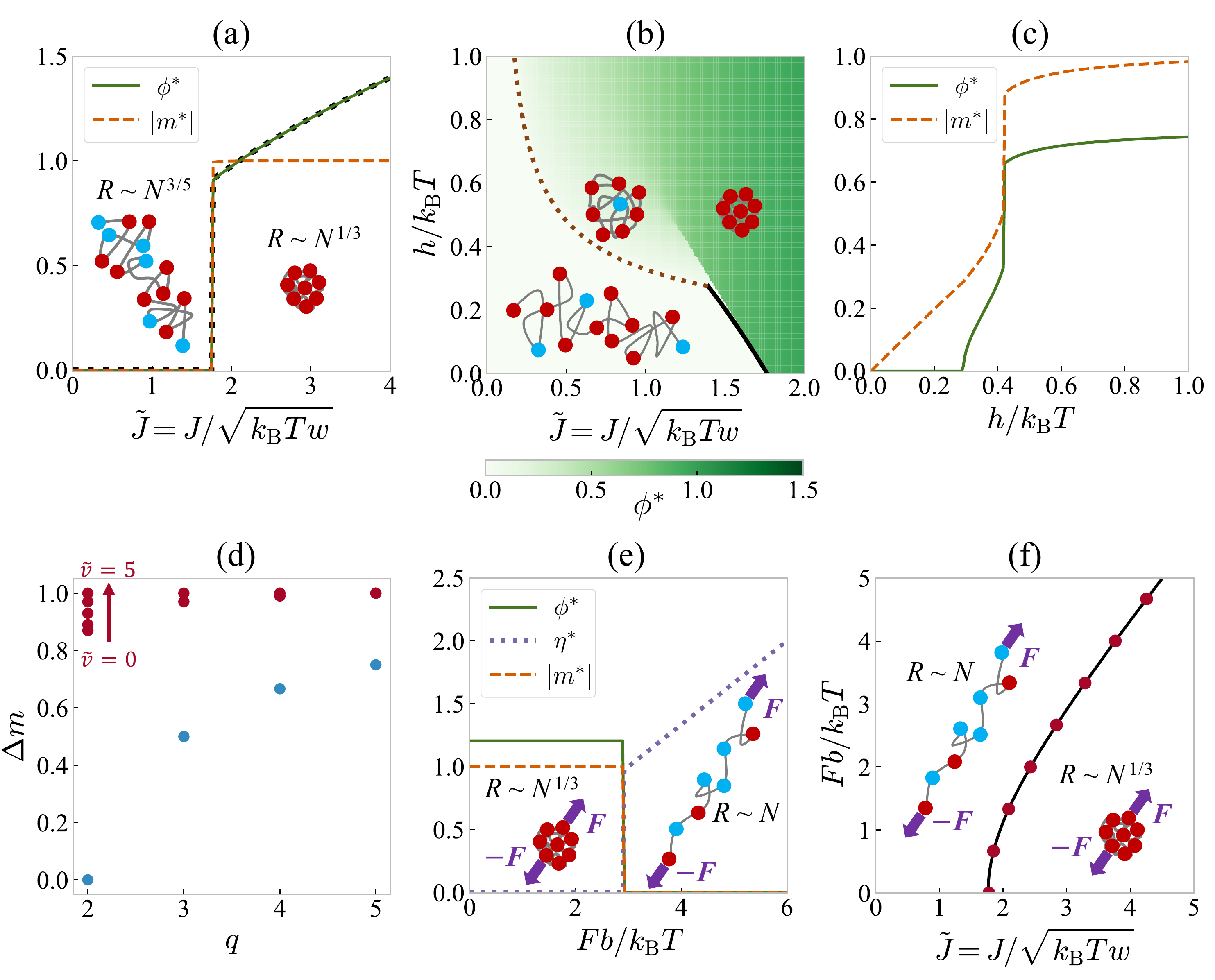}
\caption{(a)~The globular order parameter $\phi^*$ (green solid line) and magnetization $m^*$ (orange dashed line) as a function of the mark-specific interaction strength $\tilde{J}$.
The approximate functional form of $\phi^*$ [Eq.~\eqref{Eq:ApproxOrderParams}] is plotted as black dotted line.
(b)~Phase diagram in the $J$-$h$ plane.
The analytical expressions of the second-order transition line~\cite{SM} (brown dotted line) and the approximate first-order transition line~\cite{SM} (black solid line) are also plotted.
(c)~The order parameters, $\phi^*$ and $m^*$, as a function of the external field strength $h$ for the case of $\tilde{J} = 1.25$.
(d)~Magnetization jump ($\Delta m$) at the magnetic transition point as a function of the number of mark types ($q$) for the mean-field polymer-Potts model [red (dark gray) points] and for the mean-field Potts model [blue (light gray) points].
For the polymer-Potts model, $\Delta m$ is plotted for several values of $\tilde{v}$ ($\tilde{v} = 0, 0.1, 0.5, 1, 5$ for $q = 2$; $\tilde{v} = 0, 5$ for $q \geq 3$).
(e)~The stretched-state order parameter $\eta^*$ (purple dotted line), in addition to $\phi^*$ (green solid line) and $m^*$ (orange dashed line), as a function of the external force strength $F$ for the case of $\tilde{J} = 3$.
(f)~Phase diagram in the $J$-$F$ plane.
The small-$J$ and large-$J$ regions represent the stretched-disordered and compact-ordered phases, respectively.
The phase boundary estimated with Eqs.~\eqref{Eq:PseudoFreeEnergyDensityPhi}, \eqref{Eq:FreeEnergyDensity_Force}, and \eqref{Eq:PseudoFreeEnergyDensity_Force} is shown with red circles, along with the approximate $F^\mathrm{(c)} (\tilde{J})$ curve~\cite{SM} (black line).
The transition is always of first-order.}
\label{Fig:PhaseDiagrams}
\end{figure}

In Fig.~\ref{Fig:PhaseDiagrams}(a) the $\tilde{J}$ dependence of $\phi^*$ and $m^*$ is plotted for $\tilde{v} = 0.1$, showing a clear jump of the order parameters at the transition point.
Since $|m^*| \simeq 1$ in the compact-ordered state (Fig.~\ref{Fig:PseudoFreeEnergy}), we can approximate $\min_{\phi, m} f (\phi, m) \simeq \min_\phi f(\phi, m = \pm 1) = -(\tilde{J} - \tilde{v})^2 k_\mathrm{B} T / 4$. We then obtain
\begin{empheq}[left=\empheqlbrace]{align}
& \tilde{J}^\mathrm{(c)} \simeq \tilde{v} + 2 \sqrt{\ln 2}
\label{Eq:ApproxCriticalJ} \\
& \phi^* \simeq \sqrt{(\tilde{J} - \tilde{v}) / 2} \quad \left( \mathrm{for} \ \tilde{J} > \tilde{J}^\mathrm{(c)} \right)
\label{Eq:ApproxOrderParams}
\end{empheq}
shown as the black dotted line in Fig.~\ref{Fig:PhaseDiagrams}(a), giving a good approximation.
We have confirmed that the features such as the jump of $|m^*|$ from $0$ to $\simeq \! \! 1$ at the transition point are observed for a broad range of the values of $\tilde{v}$~\cite{SM}.

We further consider the effect of external mark-specific fields by setting $h_1 = h$ and $h_2 = - h$ in the Ising-type model.
The phase diagram in the case of $\tilde{v} = 0.1$ is shown in Fig.~\ref{Fig:PhaseDiagrams}(b).
We find that the field-induced transition from the swollen-disordered state to the compact-ordered state is of first-order around the zero-field transition point, meaning that increasing or decreasing specific writers can also induce the switching behavior.
Interestingly, within a certain range of $\tilde{J}$, sequential second- and first-order transitions occur as the field becomes stronger [Fig.~\ref{Fig:PhaseDiagrams}(c)].
For a smaller $\tilde{J}$, a single continuous transition is induced by the field.
Note that such a field-induced transition has been discussed in the context of the magnetic polymer within mean-field approaches~\cite{Garel1999} as well as in simulations on self-avoiding walk models~\cite{Garel1999, Huang2004}.

\textit{Propeties of transitions under general settings.---}
Here we consider the condition for the transition between the swollen and compact states to be countinuous under general $q$, $\{ J_{ij} \}$, and $\{ h_i \}$.
Minimizing Eq.~\eqref{Eq:PseudoFreeEnergyDensityPhi} on the assumption that the continuous transition occurs at $\{ \tilde{J}_{ij}^\mathrm{(c)} \}$ for a given set of $\{ h_i \}$, we obtain $\phi^* = 0$ and $x_i^* = S_i (\tilde{\bm{h}}) := \exp (h_i / k_\mathrm{B} T) / \sum_{j=1}^q \exp (h_j / k_\mathrm{B} T)$~\cite{SM}.
The order parameters will grow in response to the deviation of the interaction strengths from their critical values: $\phi^* = \Delta \phi$ and $x_i^* = S_i (\tilde{\bm{h}}) + \Delta x_i$ for $\tilde{J}_{ij} = \tilde{J}_{ij}^\mathrm{(c)} + \Delta \tilde{J}_{ij}$.
Minimizing Eq.~\eqref{Eq:PseudoFreeEnergyDensityPhi} at $\tilde{J}_{ij} = \tilde{J}_{ij}^\mathrm{(c)} + \Delta \tilde{J}_{ij}$ will give the following relation:
\begin{equation}
2 \left( \Delta \phi \right)^2 = \sum_{1 \leq i, j \leq q} \left( \tilde{J}_{ij}^\mathrm{(c)} + \Delta \tilde{J}_{ij} \right) \left[ S_i \left( \tilde{\bm{h}} \right) + \Delta x_i \right] \left[ S_j \left( \tilde{\bm{h}} \right) + \Delta x_j \right] - \tilde{v}.
\end{equation}
For the transition to be continuous, the order parameters should smoothly change at the transition point: $\Delta \phi \to 0$ and $\Delta x_i \to 0$ for $\Delta \tilde{J}_{ij} \to 0$, meaning that $J_{ij}^\mathrm{(c)}$ should obey
\begin{equation}
\sum_{1 \leq i, j \leq q} J_{ij}^\mathrm{(c)} S_i \left( \tilde{\bm{h}} \right) S_j \left( \tilde{\bm{h}} \right) = v.
\end{equation}
Therefore, if $\sum_{i,j} J_{ij} S_i (\tilde{\bm{h}}) S_j (\tilde{\bm{h}}) < v$ is satisfied, which is when the mean effective two-body interaction is repulsive, any continuous swollen-compact transition is prohibited and only switch-like transitions can occur.
This condition can in principle be checked in experiment by measuring the effective interactions between nucleosomes.

A simple example is again the Ising-type model without external fields.
In this case, since $\sum_{i,j} J_{ij} S_i (\tilde{\bm{h}}) S_j (\tilde{\bm{h}}) =  \sum_{i,j} J_{ij}/4 = 0 < v$, the continuous conformation transition is always prohibited, consistent with our numerical results that the first-order transition occurs irrespective of the value of $v$~\cite{SM}.

To investigate the order-parameter jump at the first-order transition more specifically, we consider the Potts-type interaction: $J_{ij} = J \delta_{ij}$ with $h_i=0$.
Figure~\ref{Fig:PhaseDiagrams}(d) shows the magnetization jump $\Delta m = \max \{ x_i^*\} - \min \{ x_i^* \}$ for $q \geq 2$ in the mean-field polymer-Potts model [Eq.~\eqref{Eq:PseudoFreeEnergyDensityPhi}] compared with the conventional mean-field Potts model.
In the Potts model~\cite{Wu1982}, the transition is of second-order for $q = 2$ while first-order for $q \geq 3$, and $\Delta m$ is given as $(q - 2) / (q - 1)$.
In the polymer-Potts model, on the other hand, the magnetic transition is always first-order, and $\Delta m$ is a monotonically increasing function of $v$ and $q$, while the dependency on $v$ is almost negligible for $q \geq 4$.
Notice that $\Delta m$ in the polymer-Potts model is always larger than that in the corresponding Potts model, and $\Delta m$ is practically unity for $q \geq 4$.
This suggests that the polymer conformation change that accompanies the magnetic transition reinforces the all-or-none switching property.

\textit{Mechanical discontinuous transition.---}
We here consider what happens when a stretching force is applied to the edges of a polymer chain with mark degrees of freedom.
For simplicity, let us investigate the effects of an external force term added to the pseudo free energy [Eq.~\eqref{Eq:PseudoFreeEnergyDensityPhi}] with the Ising-type interaction.
The force term can be written as $f_\mathrm{F} = - \bm{F} \cdot \bm{R} / N$ with an external force $\bm{F}$ and the polymer end-to-end vector $\bm{R}$.

Within the mean-field level, the free energy including the effects of an external force is given as
\begin{equation}
\overline{f} = \min \ \{ \min_{\phi, \, m} f(\phi, m), \min_{\eta, \, m} f'(\eta, m) \},
\label{Eq:FreeEnergyDensity_Force}
\end{equation}
where $f' (\eta, m)$ is another pseudo free energy including the effect of the external force:
\begin{eqnarray}
\frac{f' (\eta, m)}{k_\mathrm{B} T} &=& - \frac{F b \eta}{k_\mathrm{B} T} + \frac{3}{2} \eta^2 \nonumber \\
&& + \frac{1 + m}{2} \ln \frac{1 + m}{2} + \frac{1 - m}{2} \ln \frac{1 - m}{2}.
\label{Eq:PseudoFreeEnergyDensity_Force}
\end{eqnarray}
Here, the dimensionless polymer length, $\eta := R / N b$, can be interpreted as an order parameter characterizing a stretched state with the scaling $R \sim N$~\cite{Lai1995}.
In Eq.~\eqref{Eq:PseudoFreeEnergyDensity_Force}, the second term represents the entropic elasticity~\cite{DeGennes1979}, which is essential under stretched conditions.
We denote the equilibrium point as ($\phi^*, \eta^*, m^*$).
Since the globule state ($\phi^* > 0$, $R \sim N^{1/3}$) and the stretched state ($\eta^* > 0$, $R \sim N$) are incompatible, only one of ($\phi^*, \eta^*$) can be finite and the other should be zero.

Figure~\ref{Fig:PhaseDiagrams}(e) shows the changes in order parameters upon varying of the external force $F$ for the case of $\tilde{v} = 0.1$ and $\tilde{J} = 3$, in which the compact-ordered state ($\phi^* > 0$, $\eta^* = 0$, and $|m^*| > 0$) is stabilized when
$F = 0$.
We can see that a first-order transition occurs at a certain critical value $F^\mathrm{(c)}$, above which a stretched-disordered state ($\phi^* = 0$, $\eta^* > 0$, and $m^* = 0$) emerges.
The numerically obtained phase diagram in the $J$-$F$ plane is shown in Fig.~\ref{Fig:PhaseDiagrams}(f).
Note that the force-induced coil-globule transitions are believed to be discontinuous also in classical polymer models at $N \to \infty$~\cite{Halperin1991, Lai1995, Grassberger2002, Geissler2002}.
In the polymer-Potts model, we find that the mark degrees of freedom become immediately disordered accompanying this stretching transition.
Similar discontinuous transitions between a compact-ordered state and a stretched-disordered state have recently been seen in molecular dynamics simulations with short-range interactions~\cite{Michieletto2017}.

\begin{figure}[t]
\includegraphics[scale=0.37]{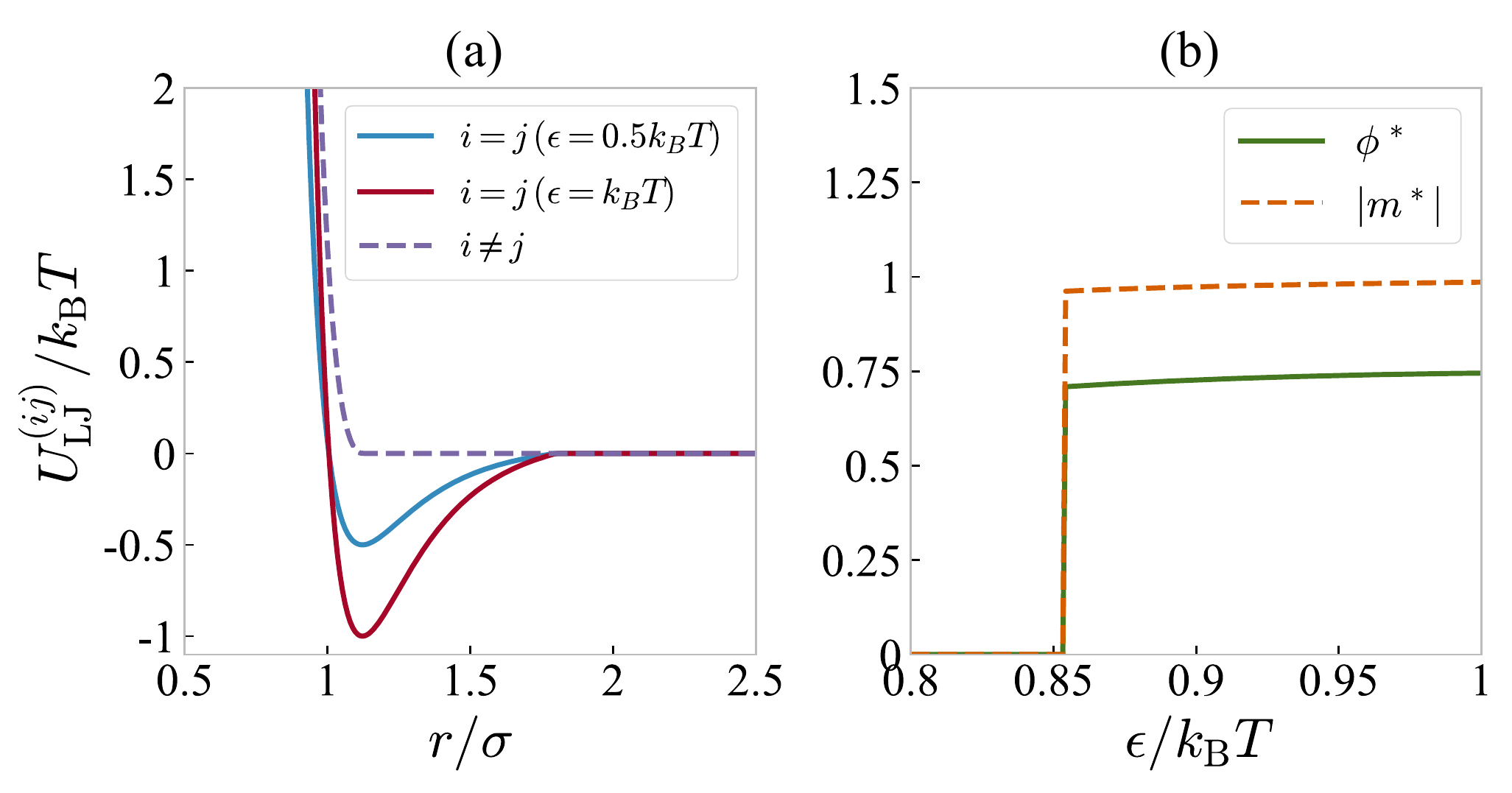}
\caption{(a)~Lennard-Jones-type interaction $U_\mathrm{LJ}^{(ij)} (r)$ between two monomers ($1 \leq i, j \leq 2$).
$\sigma$ is the interaction length, and $\epsilon$ is the attractive interaction strength between monomers with the same mark.
(b)~Optimized order parameters $\phi^*$ and $m^*$ as a function of the interaction strength $\epsilon / k_\mathrm{B} T$.}
\label{Fig:OrderParamsVirial}
\end{figure}

\textit{Relation to molecular dynamics simulations.---}
To compare our results with the molecular dynamics simulations~\cite{Michieletto2016} using Lennard-Jones-type interactions [Fig.~\ref{Fig:OrderParamsVirial}(a)], we consider the virial expansion.
By neglecting $\mathcal{O} (\phi^6)$ terms and the existence of the neutral mark, we obtain the pseudo free energy [Eq.~\eqref{Eq:PseudoFreeEnergyDensityPhi}] with additional terms proportional to $m^2 \phi^4$~\cite{SM}.

By minimizing the pseudo free energy, we obtain the optimized $\phi^*$ and $m^*$ as a function of the interaction strength $\epsilon / k_\mathrm{B} T$.
Figure~\ref{Fig:OrderParamsVirial}(b) shows that a discontinuous transition with a large jump of magnetization occurs at the interaction strength $\epsilon / k_\mathrm{B} T \simeq 0.85$, which is close to the simulation result~\cite{Michieletto2016} ($\epsilon / k_\mathrm{B} T \simeq 0.9$ for $N = 2000$).
Although the virial expansion is not generally justfied for cases with a first-order transition, this result suggests that a simplified framework can connect the molecular level measurement of histone interactions~\cite{Funke2016} to the compartment level chromatin state transition.

\textit{Discussion and conclusion.---}
Here we have studied the polymer-Potts model at the mean-field level and found that switch-like transitions are largely enhanced, compared with the transitions in a polymer model with unchangeable marks or the conventional Potts model, due to the coordination of the coil-globule and magnetic transition.
The bistable property leading to the first-order transition fits with the phenomenology of chromatin state transition and cell differentiation.
For instance, it has been shown that elimination of small kilobase-scale genome regions can induce compartment switching of a whole megabase-scale region~\cite{Sima2019}.
This can be explained by the bistability of the chromatin state, which allows localized histone mark biases induced by transcription factors to spread macroscopically.
The hysteresis effect, which is expected to accompany the chromatin discontinuous transition, may also improve the stability of the epigenetic regulation against chemical and mechanical perturbations and cell division.

Additional to the equilibrium phase transition scenario proposed in this paper, nonequilibrium features of the chemical dynamics~\cite{Michieletto2016,Coli2018} and the phase separation properties of the key components in chromatin dynamics~\cite{Gibson2019, Strom2017, Larson2017, Plys2018, Tatavosian2019} may play roles in enhancing or diminishing the switch-like behavior.
Nevertheless, the fact that a simple mark-conformation coupling can lead to a discrete switch indicates that nonlinear dynamics and well-designed chemical networks may not be essential in explaining cell fate dynamics.  
In real differentiation, state switching occurs in sub-regions and does not expand to the whole chromosome~\cite{Dixon2015}.
It is interesting to explore how specific regions in the genome set boundaries to prevent the phase transition dynamics from spreading into undesired regions~\cite{Obersriebnig2016}.

\textit{Acknowledgements.---}
We are grateful to Yohsuke T. Fukai and Soya Shinkai for fruitful discussions and the reading of the manuscript.
This work was supported by JSPS KAKENHI Grant Numbers JP18H04760, JP18K13515, JP19H05275, JP19H05795.


%

\pagebreak

\onecolumngrid

\setcounter{equation}{0}
\setcounter{figure}{0}
\setcounter{table}{0}
\setcounter{page}{1}
\renewcommand{\thepage}{S\arabic{page}}  
\renewcommand{\thetable}{S\arabic{table}}   
\renewcommand{\thefigure}{S\arabic{figure}}
\renewcommand{\theequation}{S\arabic{equation}}

\begin {center} 
	\textbf{ \large Supplemental Material for \\ Chromatin state switching in a model with mark-conformation coupling}\\ [.1cm]
	
	{Kyosuke Adachi and Kyogo Kawaguchi}, \\ [.1cm]
	{(Dated: \today)} \\
\end {center}

\section{polymer-Potts model}
\label{Sec:PolymerPotts}

In this section, we describe how to obtain the pseudo free energy of the polymer-Potts model starting a generic form of microscopic interactions.
The Hamiltonian of the polymer-Potts model with external field and force terms reads
\begin{equation}
H = H_\mathrm{GC} (\{ {\bm R}_n \}) + H_\mathrm{P} (\{ {\bm R}_n \}, \{ s_n \}) + H_\mathrm{M} (\{ s_n \}) + H_\mathrm{VE} (\{ {\bm R}_n \}) + H_\mathrm{F} (\bm{R}_0, \bm{R}_N).
\label{Eq:PolymerPotts}
\end{equation}
Here, the degrees of freedom are the marks of each monomer, $s_n$ $(1 \leq n \leq N)$, which represent the types of chemical modifications attached to the $n$-th histone along the chain, and the positions of the monomers, $\bm{R}_n$.
We assume that there are $q$ $(\geq 2)$ types of histone marks that are distinguishable in the sense of interactions: $s_n \in \{ 1, 2, ..., q \}$.
Also, we take $\{ \bm{R}_n \}_{n=1}^N$ as a set of variables.
The first term of Eq.~(\ref{Eq:PolymerPotts}),
\begin{equation}
H_\mathrm{GC} (\{ {\bm R}_n\})= \frac{3 k_\mathrm{B} T}{2 b^2} \sum_{n = 1}^N \left( \bm{R}_n - \bm{R}_{n - 1} \right)^2,
\label{Eq:GaussianChain}
\end{equation}
corresponds to the Gaussian-chain interaction that restricts the positions of monomers adjacent to each other in the polymer, where $b$ corresponds to the Kuhn length (the size of the monomer), and $T$ is the temperature.
The second term,
\begin{equation}
H_\mathrm{P} (\{ {\bm R}_n\}, \{ s_n\}) = - \sum_{n \neq m} J(s_n, s_m) \delta \left( \bm{R}_n - \bm{R}_m \right),
\end{equation}
describes the attractive or repulsive interaction between the monomers when they become close in contact with each other.
The detail of interactions between the different types of marks can be coded in $J (s_n, s_m) = \sum_{1 \leq i, j \leq q} J_{ij} \delta_{i, s_n} \delta_{j, s_m}$, where $J_{ij}$ is a real symmetric matrix.
For example, the Ising-type interaction can be realized if we set $q = 2$ and $J_{ij} = J ( 2\delta_{ij} -1)$, and the $q$-state Potts-type interaction can be realized if we choose $J_{ij} = J \delta_{i j}$.
The third term,
\begin{equation}
H_\mathrm{M} (\{ s_n \}) = - \sum_{i} h_i \sum_n \delta_{s_n, i},
\end{equation}
describes the effect of an external bias field, which favors a certain type of histone marks over other types.
The fourth term,
\begin{equation}
H_\mathrm{VE}(\{ {\bm R}_n\}) = v \sum_{n \neq m} \delta \left( \bm{R}_n - \bm{R}_m \right) + w \sum_{n \neq m \neq k} \delta \left( \bm{R}_n - \bm{R}_m \right) \delta \left( \bm{R}_m - \bm{R}_k \right),
\end{equation}
describes the two- and three-body volume exclusion interaction terms that we assume have no histone-mark dependence ($v, w > 0$).
The last term,
\begin{equation}
H_\mathrm{F} (\bm{R}_0, \bm{R}_N) := - \bm{F} \cdot (\bm{R}_N - \bm{R}_0),
\end{equation}
represents a mechanical stretching force that is applied to the edges of the polymer. 

\subsection{Modified polymer-Potts model with long-range interactions}

To perform a mean-field analysis of Eq.~\eqref{Eq:PolymerPotts}, let us replace the local quantities with the averaged counterparts.
First, we rewrite the interaction terms ($H_\mathrm{P}$ and $H_\mathrm{VE}$) with the polymer density, $\rho (\bm{r}, \{ \bm{R}_n \}) := \sum_n \delta (\bm{r} - \bm{R}_n)$, and the mark density of the $i$-th state, $\sigma_i (\bm{r}, \{ \bm{R}_n \}, \{ s_n \}) := \sum_n \delta_{i, s_n} \delta (\bm{r} - \bm{R}_n)$.
We then obtain 
\begin{equation}
H_\mathrm{P} (\{ {\bm R}_n\}, \{ s_n\}) = - \sum_{i, j} J_{i j} \int \mathrm{d} \bm{r} \sigma_i (\bm{r}, \{ \bm{R}_n \}, \{ s_n \}) \sigma_j (\bm{r}, \{ \bm{R}_n \}, \{ s_n \}),
\end{equation}
and
\begin{equation}
H_\mathrm{VE} (\{ {\bm R}_n\}) = v \int \mathrm{d} \bm{r} \rho (\bm{r}, \{ \bm{R}_n \})^2 + w \int \mathrm{d} \bm{r} \rho (\bm{r}, \{ \bm{R}_n \})^3,
\end{equation}
by neglecting the terms proportional to $\delta (\bm{0})$.
Next, we define the average density and mark density as
\begin{equation}
\rho_\mathrm{av} (\bm{r}, \bm{R}_{N0}) := \frac{3 N}{4 \pi {R_{N0}}^3} \theta (R_{N0} - r),
\end{equation}
and
\begin{equation}
\sigma_{i, \mathrm{av}} (\bm{r}, \bm{R}_{N0}, \{ s_n \}) := \rho_\mathrm{av} (\bm{r}, \bm{R}_{N0}) \ \frac{1}{N} \sum_n \delta_{i, s_n},
\end{equation}
where $\bm{R}_{N0} := \bm{R}_N - \bm{R}_0$ is the end-to-end distance vector and $\theta (x)$ is the Heaviside step function.
Note that $\int \mathrm{d} \bm{r} \rho_\mathrm{av} (\bm{r}, \bm{R}_{N0}) = \int \mathrm{d} \bm{r} \rho (\bm{r}, \{ \bm{R}_n \}) = N$ and $\int \mathrm{d} \bm{r} \sigma_{i, \mathrm{av}} (\bm{r}, \bm{R}_{N0}, \{ s_n \}) = \int \mathrm{d} \bm{r} \sigma_{i} (\bm{r}, \{ \bm{R}_n \}, \{ s_n \}) = \sum_n \delta_{i, s_n}$.

As a mean-field approximation, we replace $\rho$ and $\sigma_i$ with $\rho_\mathrm{av}$ and $\sigma_{i, \mathrm{av}}$, respectively.
The obtained Hamiltonian is a modified polymer-Potts model with long-range interactions:
\begin{equation}
H^\mathrm{LR} = H_\mathrm{GC} (\{ {\bm R}_n \}) + H_\mathrm{P}^\mathrm{LR} (\bm{R}_{N0}, \{ s_n \}) + H_\mathrm{M} (\{ s_n \}) + H_\mathrm{VE}^\mathrm{LR} (\bm{R}_{N0}) + H_\mathrm{F} (\bm{R}_0, \bm{R}_N).
\label{Eq:PolymerPottsLongRange}
\end{equation}
The long-range interaction terms ($H_\mathrm{P}^\mathrm{LR}$ and $H_\mathrm{VE}^\mathrm{LR}$) are given as
\begin{eqnarray}
H_\mathrm{P}^\mathrm{LR} (\bm{R}_{N0}, \{ s_n \}) &:=& - \sum_{i, j} J_{i j} \int \mathrm{d} \bm{r} \sigma_{i, \mathrm{av}} (\bm{r}, \bm{R}_{N0}, \{ s_n \}) \sigma_{j, \mathrm{av}} (\bm{r}, \bm{R}_{N0}, \{ s_n \}) \nonumber \\
&=& - \frac{N^2}{{R_{N0}}^3} \sum_{i, j} J'_{i j} \left( \frac{1}{N} \sum_n \delta_{i, s_n} \right) \left( \frac{1}{N} \sum_m  \delta_{j, s_m} \right),
\end{eqnarray}
where $J'_{ij} := 3 J_{i j} / 4 \pi$, and
\begin{eqnarray}
H_\mathrm{VE}^\mathrm{LR} (\bm{R}_{N0}) &:=& v \int \mathrm{d} \bm{r} \rho_\mathrm{av} (\bm{r}, \bm{R}_{N0})^2 + w \int \mathrm{d} \bm{r} \rho_\mathrm{av} (\bm{r}, \bm{R}_{N0})^3 \nonumber \\
&=& v' \frac{N^2}{{R_{N0}}^3} + w' \frac{N^3}{{R_{N0}}^6},
\label{Eq:VolumeExclusionLR}
\end{eqnarray}
where $v' := 3 v / 4 \pi$ and $w' := (3 / 4 \pi)^2 w$.

\subsection{Derivation of pseudo free energy}

In the following, we derive the functional form of the pseudo free energy (per monomer) through analyzing the equilibrium free energy of the modified polymer-Potts model, $\overline{f} := -(k_\mathrm{B} T / N) \ln Z$, where the canonical partition function is defined as
\begin{equation}
Z := \left( \prod_{n = 1}^N \int \mathrm{d} \bm{R}_n \sum_{s_n = 1}^q \right) \exp \left( -\frac{H^\mathrm{LR}}{k_\mathrm{B} T} \right).
\end{equation}
Using the explicit form of $H^\mathrm{LR}$ and an identity, $\int \mathrm{d} \bm{R} \, \delta (\bm{R} - \bm{R}_{N0}) = 1$, we can obtain
\begin{eqnarray}
Z &=& \int \mathrm{d} \bm{R} \exp \left( - \frac{v'}{k_\mathrm{B} T} \frac{N^2}{R^3} - \frac{w'}{k_\mathrm{B} T} \frac{N^3}{R^6} + \frac{\bm{F} \cdot \bm{R}}{k_\mathrm{B} T} \right) \ \left( \prod_{n = 1}^N \sum_{s_n = 1}^q \right) \exp \left( \frac{1}{R^3} \sum_{i, j} \frac{J'_{ij}}{k_\mathrm{B} T} \sum_{n, m} \delta_{i, s_n} \delta_{j, s_m} + \sum_{i = 1}^q \frac{h_i}{k_\mathrm{B} T} \sum_{n = 1}^N \delta_{i, s_n} \right) \nonumber \\
&& \times \left( \prod_{n = 1}^N \int \mathrm{d} \bm{R}_n \right) \delta (\bm{R} - \bm{R}_{N0}) \exp \left[ - \frac{3}{2 b^2} \sum_{n = 1}^N (\bm{R}_n - \bm{R}_{n - 1})^2 \right].
\label{Eq:CalculationIdentity}
\end{eqnarray}
Note that $\bm{R}$ represents the end-to-end distance vector, which is clear from its definition.

To perform the integration with respect to $\{ \bm{R}_n \}_{n = 1}^N$, we change the integration variables from $\{ \bm{R}_n \}_{n = 1}^N$ to $\{ \overline{\bm{R}}_n := \bm{R}_n - \bm{R}_{n - 1} \}_{n = 1}^N$.
Using $\delta (\bm{R} - \bm{R}_{N0}) = \delta (\bm{R} - \sum_{n = 1}^N \overline{\bm{R}}_n) = (2 \pi)^{-3} \int \mathrm{d} \bm{k} \exp [\mathrm{i} \bm{k} \cdot (\bm{R} - \sum_{n = 1}^N \overline{\bm{R}}_n)]$, we can proceed the calculation as
\begin{eqnarray}
&& \left( \prod_{n = 1}^N \int \mathrm{d} \bm{R}_n \right) \delta (\bm{R} - \bm{R}_{N0}) \exp \left[ - \frac{3}{2 b^2} \sum_{n = 1}^N (\bm{R}_n - \bm{R}_{n - 1})^2 \right] \nonumber \\
&& = \int \frac{\mathrm{d} \bm{k}}{(2 \pi)^3} \exp \left( \mathrm{i} \bm{k} \cdot \bm{R} \right) \left( \prod_{n = 1}^N \int \mathrm{d} \overline{\bm{R}}_n \right) \exp \left( - \frac{3}{2 b^2} \sum_{n = 1}^N {\overline{\bm{R}}_n}^2 - \mathrm{i} \bm{k} \cdot \sum_{n = 1}^N \overline{\bm{R}}_n \right) \nonumber \\
&& = \left( \frac{2 \pi b^2}{3} \right)^{3N / 2} \int \frac{\mathrm{d} \bm{k}}{(2 \pi)^3} \exp \left( - \frac{N b^2 {\bm{k}}^2}{6} + \mathrm{i} \bm{k} \cdot \bm{R} \right) \nonumber \\
&& = \frac{1}{N^{3/2}} \left( \frac{2 \pi b^2}{3} \right)^{3(N - 1) / 2} \exp \left( - \frac{3 R^2}{2 N b^2} \right).
\label{Eq:CalculationGC}
\end{eqnarray}
We then obtain
\begin{eqnarray}
Z &=& \frac{1}{N^{3/2}} \left( \frac{2 \pi b^2}{3} \right)^{3(N - 1) / 2} \int \mathrm{d} \bm{R} \exp \left( - \frac{3 R^2}{2 N b^2} - \frac{v'}{k_\mathrm{B} T} \frac{N^2}{R^3} - \frac{w'}{k_\mathrm{B} T} \frac{N^3}{R^6} + \frac{\bm{F} \cdot \bm{R}}{k_\mathrm{B} T} \right) \nonumber \\
&& \times \left( \prod_{n = 1}^N \sum_{s_n = 1}^q \right) \exp \left( \frac{1}{R^3} \sum_{i, j} \frac{J'_{ij}}{k_\mathrm{B} T} \sum_{n, m} \delta_{i, s_n} \delta_{j, s_m} + \sum_{i = 1}^q \frac{h_i}{k_\mathrm{B} T} \sum_{n = 1}^N \delta_{i, s_n} \right).
\end{eqnarray}

To perform the summation with respect to $\{ s_n \}_{n = 1}^N$, we use the identity, $\sum_{N_i = 0}^N \delta (N_i, \sum_{n = 1}^N \delta_{i, s_n}) = 1$, where $\delta (a, b) := \delta_{a, b}$ is the Kronecker delta.
We then obtain
\begin{eqnarray}
&& \left( \prod_{n = 1}^N \sum_{s_n = 1}^q \right) \exp \left( \frac{1}{R^3} \sum_{i, j} \frac{J'_{ij}}{k_\mathrm{B} T} \sum_{n, m} \delta_{i, s_n} \delta_{j, s_m} + \sum_{i = 1}^q \frac{h_i}{k_\mathrm{B} T} \sum_{n = 1}^N \delta_{i, s_n} \right) \nonumber \\
&& = \left( \prod_{i = 1}^q \sum_{N_i = 0}^N \right) \exp \left( \frac{1}{R^3} \sum_{i, j} \frac{J'_{ij}}{k_\mathrm{B} T} N_i N_j + \sum_{i = 1}^q \frac{h_i}{k_\mathrm{B} T} N_i \right) \left( \prod_{n = 1}^N \sum_{s_n = 1}^q \right) \left[ \prod_{i = 1}^q \delta \left( N_i, \sum_{n = 1}^N \delta_{i, s_n} \right) \right] \nonumber \\
&& = \left( \prod_{i = 1}^q \sum_{N_i = 0}^N \right) \exp \left( \frac{1}{R^3} \sum_{i, j} \frac{J'_{ij}}{k_\mathrm{B} T} N_i N_j + \sum_{i = 1}^q \frac{h_i}{k_\mathrm{B} T} N_i \right) \frac{N!}{\prod_{i = 1}^q N_i !} \, \delta \left( \sum_{i = 1}^q N_i, N \right).
\end{eqnarray}
Note that $N_i$ represents the occupation number of the $i$-th mark state.
Applying Stirling's formula, we can estimate the terms originated from the entropy as
\begin{equation}
\frac{N!}{\prod_{i = 1}^q N_i !} = \exp \left( - \sum_{i = 1}^q N_i \ln \frac{N_i}{N} - \frac{1}{2} (q - 1) \ln N + O(N^0) \right).
\end{equation}
Replacing the summation with respect to $N_i$ with an integration over the occupation ratio of the $i$-th mark state, $x_i := N_i / N$, we obtain
\begin{eqnarray}
&& \left( \prod_{i = 1}^q \sum_{N_i = 0}^N \right) \exp \left( \frac{1}{R^3} \sum_{i, j} \frac{J'_{ij}}{k_\mathrm{B} T} N_i N_j + \sum_{i = 1}^q \frac{h_i}{k_\mathrm{B} T} N_i \right) \frac{N!}{\prod_{i = 1}^q N_i !} \, \delta \left( \sum_{i = 1}^q N_i, N \right) \nonumber \\
&& = N^{q -1} \left( \prod_{i = 1}^q \int_0^1 \mathrm{d} x_i \right) \delta \left( \sum_{i = 1}^q x_i - 1 \right) \exp \left[ \frac{N^2}{R^3} \sum_{i, j} \frac{J'_{ij}}{k_\mathrm{B} T} x_i x_j + N \sum_{i = 1}^q \frac{h_i}{k_\mathrm{B} T} x_i - N \sum_{i = 1}^q x_i \ln x_i - \frac{q - 1}{2} \ln N + O(N^0) \right] \nonumber \\
&& = \left( \prod_{i = 1}^q \int_0^1 \mathrm{d} x_i \right) \delta \left( \sum_{i = 1}^q x_i - 1 \right) \exp \left[ \frac{N^2}{R^3} \sum_{i, j} \frac{J'_{ij}}{k_\mathrm{B} T} x_i x_j + N \sum_{i = 1}^q \frac{h_i}{k_\mathrm{B} T} x_i - N \sum_{i = 1}^q x_i \ln x_i  + \frac{q - 1}{2} \ln N + O (N^0) \right].
\end{eqnarray}

Based on the above calculation, after performing the solid-angle integration of $\bm{R}$, we finally obtain the following reduced form of the partition function:
\begin{equation}
Z = \int_0^\infty \mathrm{d} R \left( \prod_{i = 1}^q \int_0^1 \mathrm{d} x_i \right) \delta \left( \sum_{i = 1}^q x_i - 1 \right) \exp \left( - \frac{N f (R, \{ x_i \} )}{k_\mathrm{B} T} - \frac{Nf_0}{k_\mathrm{B} T} + O (N^0) \right),
\label{Eq:PartitionFunctionFinal} 
\end{equation}
where the constant parts of the free energy, $f_0$, is defined as
\begin{equation}
\frac{f_0}{k_\mathrm{B} T} := \frac{3}{2} \ln \left( \frac{2 \pi b^2}{3} \right) - \frac{q - 2}{2} \frac{\ln N}{N},
\end{equation}
and the pseudo free energy, $f (R, \{ x_i \})$, is given as
\begin{equation}
\frac{f (R,  \{ x_i \})}{k_\mathrm{B} T} := \frac{3 R^2}{2 N^2 b^2} - \frac{F R}{k_\mathrm{B} T N} + \frac{v' - \sum_{i, j} J'_{ij} x_i x_j}{k_\mathrm{B} T} \frac{N}{R^3} + \frac{w'}{k_\mathrm{B} T} \frac{N^2}{R^6} - \sum_{i = 1}^q \frac{h_i}{k_\mathrm{B} T} x_i + \sum_{i = 1}^q x_i \ln x_i - \frac{2}{N} \ln \frac{R}{\sqrt{N} b}.
\label{Eq:PseudoFreeEnergyFiniteSize}
\end{equation}
In the limit of $N \to \infty$, the last term in Eq.~\eqref{Eq:PseudoFreeEnergyFiniteSize} is negligible, and we can apply  Laplace's method to estimate Eq.~\eqref{Eq:PartitionFunctionFinal}.
Thus, the free energy, $\overline{f} = -(k_\mathrm{B} T / N) \ln Z$, is obtained as
\begin{equation}
\lim_{N \to \infty} \overline{f} = \lim_{N \to \infty} \left[ \min_{R, \, \{ x_i \}} f (R, \{ x_i \}) \mathrm{\ subject \ to} \sum_{i = 1}^q x_i = 1 \right] + \lim_{N \to \infty} f_0.
\label{Eq:FreeEnergyDensity}
\end{equation}

Now we introduce two kinds of dimensionless order parameters, $\phi$ and $\eta$, regarding the conformation of the polymer.
The first parameter, $\phi$, is defined as
\begin{equation}
\phi := \sqrt{\frac{N}{R^3} \sqrt{\frac{w'}{k_\mathrm{B} T}}},
\end{equation}
which corresponds to the square root of the globule density ($N / R^3$) and becomes finite in the compact (or globule) state satisfying $R \sim N^{1 / 3}$.
The second parameter, $\eta$, is defined as
\begin{equation}
\eta := \frac{R}{Nb},
\end{equation}
which represents how the polymer is stretched and becomes finite in the stretched state satisfying $R \sim N$.
Note that either of $\phi$ or $\eta$ can be finite in the limit of $N \to \infty$, which can be seen from the definitions of these order parameters.
In the following, the order parameters that minimize the pseudo free energy (i.e., equilibrium state) are denoted as $\phi^*$ and $\eta^*$.
Considering the scaling $R \sim N^\nu$, the swollen state  ($\nu \simeq 3 / 5$~\cite{DeGennes1979, Doi1986}) is included in the phase characterized by $\phi^* = \eta^* = 0$.
Although our framework cannot determine $\nu$ beyond the inequality $1 / 3 < \nu < 1$ within the swollen state in the $N \to \infty$ limit, $\nu = 3/5$ is obtained in the parameter regime far from the transition point if we consider the finite size effect using Eq.~\eqref{Eq:PseudoFreeEnergyFiniteSize}~\cite{DeGennes1975}.

Regarding the macroscopic mark, the occupation ratios of each mark, $\{ x_i = N_i / N \}_{i = 1}^q$, are order parameters.
We denote the equilibrium mark order parameter as $x_i^*$, and call that the $i$-th mark is ordered when $x_i^* > 1 / q$.

Using $\phi$, $\eta$, and $\{ x_i \}$, we can rewrite the formula of the free energy [Eq.~\eqref{Eq:FreeEnergyDensity}] as
\begin{equation}
\lim_{N \to \infty} \overline{f} = \min \left\{ \overline{f}_\mathrm{C}, \, \overline{f}_\mathrm{S} \right\} + \lim_{N \to \infty} f_0.
\label{Eq:FreeEnergyDensityFinal}
\end{equation}
The compact-state free energy, $\overline{f}_\mathrm{C}$, is defined as
\begin{equation}
\overline{f}_\mathrm{C} := \left[ \min_{\phi, \, \{ x_i \}} f_\mathrm{C} (\phi, \{ x_i \}) \mathrm{\ subject \ to} \sum_{i = 1}^q x_i = 1 \right],
\label{Eq:FreeEnergyDensity_Compact}
\end{equation}
where the corresponding pseudo free energy is given as
\begin{equation}
\frac{f_\mathrm{C} (\phi, \{ x_i \})}{k_\mathrm{B} T} := \left( \tilde{v} - \sum_{i, j} \tilde{J}_{ij} x_i x_j \right) \phi^2 + \phi^4 + \frac{f_\mathrm{mark} (\{ x_i \})}{k_\mathrm{B} T}.
\label{Eq:PseudoFED_Globule}
\end{equation}
Here, $\tilde{v} := v' / \sqrt{k_\mathrm{B} T w'} = v / \sqrt{k_\mathrm{B} T w}$ and $\tilde{J}_{ij} := J'_{ij} / \sqrt{k_\mathrm{B} T w'} = J_{ij} / \sqrt{k_\mathrm{B} T w}$ are dimensionless interaction strengths, and
\begin{equation}
\frac{f_\mathrm{mark} (\{ x_i \})}{k_\mathrm{B} T} := - \sum_{i = 1}^q \tilde{h}_i x_i + \sum_{i = 1}^q x_i \ln x_i,
\end{equation}
is the non-interacting pseudo free energy of the marks with
$\tilde{h}_i := h_i / k_\mathrm{B} T$ being the dimensionless external field.
On the other hand, the stretched-state free energy, $\overline{f}_\mathrm{S}$, is defined as
\begin{equation}
\overline{f}_\mathrm{S} := \left[ \min_{\eta, \, \{ x_i \}} f_\mathrm{S} (\eta, \{ x_i \}) \mathrm{\ subject \ to} \sum_{i = 1}^q x_i = 1 \right],
\end{equation}
where the corresponding pseudo free energy is given as
\begin{equation}
\frac{f_\mathrm{S} (\eta, \{ x_i \})}{k_\mathrm{B} T} := \frac{3}{2} \eta^2  - \tilde{F} \eta  + \frac{f_\mathrm{mark} (\{ x_i \})}{k_\mathrm{B} T}.
\label{Eq:PseudoFED_Stretched}
\end{equation}
Here, $\tilde{F} := F b / k_\mathrm{B} T$ is a dimensionless stretching force strength.

To sum up, in the case of $\overline{f}_\mathrm{C} < \overline{f}_\mathrm{S}$ with $\phi^* > 0$ and $\eta^* = 0$, the compact state ($R \sim N^{1 / 3}$) is stable and the realized order parameters minimize $f_\mathrm{C} (\phi, \{ x_i \})$; in the case of $\overline{f}_\mathrm{C} > \overline{f}_\mathrm{S}$ with $\phi^* = 0$ and $\eta^* > 0$, the stretched state ($R \sim N$) is stable and the realized order parameters minimize $f_\mathrm{S} (\eta, \{ x_i \})$; in the case of $\overline{f}_\mathrm{C} = \overline{f}_\mathrm{S} = \min_{\{ x_i \}} f_\mathrm{mark} (\{ x_i \})$ with $\phi^* = \eta^* = 0$, the swollen state ($R \sim N^{3 / 5}$) is stable.
The stable polymer state is also classified into ordered and disordered states according to the occupancy of each mark: the $i$-th mark is ordered (disordered) in the case of $x_i^* > 1 /q$ ($x_i^* \leq 1 / q$).

\subsection{Derivation of equations for the order parameters}

Based on \eqref{Eq:FreeEnergyDensityFinal}, we derive the equations for the order parameters.

\subsubsection{Compact state}

To obtain the order parameters in the compact (or globule) state ($\phi^* > 0$ and $\eta^* = 0$), we need to minimize $f_\mathrm{C} (\phi, \{ x_i \})$ subject to the constraint $\sum_{i = 1}^q x_i = 1$.
Instead of directly treating the constraint, we introduce the Lagrange multiplier $\mu$.
Then, as a necessary condition, the optimized order parameters should correspond to a stationary point of the following function:
\begin{equation}
\frac{f'_\mathrm{C} (\phi, \{ x_i \}, \mu)}{k_\mathrm{B} T} := \left( \tilde{v} - \sum_{i, j} \tilde{J}_{ij} x_i x_j \right) \phi^2 + \phi^4 - \sum_{i = 1}^q \tilde{h}_i x_i + \sum_{i = 1}^q x_i \ln x_i - \mu \left( \sum_{i = 1}^q x_i - 1 \right).
\label{Eq:PseudoFED_Globule_Lagrange}
\end{equation}
We can find the stationary points of $f'_\mathrm{C} (\phi, \{ x_i \}, \mu)$ by solving $\partial f'_\mathrm{C} / \partial \phi = 0$, $\{ \partial f'_\mathrm{C} / \partial x_i = 0 \}_{i = 1}^q$, and $\partial f'_\mathrm{C} / \partial \mu = 0$, i.e.,
\begin{empheq}[left=\empheqlbrace]{align}
& \left( \tilde{v} - \sum_{i, j} \tilde{J}_{ij} x_i^* x_j^* + 2 {\phi^*}^2 \right) \phi^* = 0
\label{Eq:OP_phi} \\
& -2 {\phi^*}^2 \sum_{j = 1}^q \tilde{J}_{ij} x_j^* - \tilde{h}_i + \ln x_i^* + 1 - \mu^* = 0 \quad (1 \leq i \leq q)
\label{Eq:OP_x} \\
& \sum_{i = 1}^q x_i^* - 1 = 0,
\label{Eq:Constraint}
\end{empheq}
where $\phi^*$, $x_i^*$, and $\eta^*$ are the solution of these equations.

Since we here consider the compact state, we look for a solution with $\phi^* > 0$. Then, Eq.~\eqref{Eq:OP_phi} reduces to
\begin{equation}
{\phi^*}^2 = \frac{1}{2} \left( {\bm{x}^*}^\mathrm{T} \, \tilde{\mathsf{J}} \, \bm{x}^* - \tilde{v} \right),
\label{Eq:OP_phi_Globule}
\end{equation}
where we use matrix representations, $[\bm{x}]_i := x_i$ and $[\tilde{\mathsf{J}}]_{ij} := \tilde{J}_{ij}$.
On the other hand, Eq.~\eqref{Eq:OP_x} can be rewritten as $x_i^* = \mathrm{e}^{\, \mu^* - 1} \exp \, (2 {\phi^*}^2 \sum_{j = 1}^q \tilde{J}_{ij} x_j^* + \tilde{h}_i)$.
Substituting this expression for $x_i^*$ in Eq.~\eqref{Eq:Constraint}, we obtain $\mathrm{e}^{\, \mu^* - 1} = [ \sum_{i = 1}^q \exp \, ( 2 {\phi^*}^2 \sum_{j = 1}^q \tilde{J}_{ij}  x_j^* + \tilde{h}_i ) ]^{-1}$.
Therefore, we obtain the following expression of $x_i^*$:
\begin{equation}
x_i^* = S_i \left( 2 {\phi^*}^2 \, \tilde{\mathsf{J}} \, \bm{x}^* + \tilde{\bm{h}} \right) \quad (1 \leq i \leq q),
\label{Eq:OP_x_Globule}
\end{equation}
where $[\tilde{\bm{h}}]_i := \tilde{h}_i$, and the softmax function, $S_i \, (\bm{y})$, is defined as
\begin{equation}
S_i \, (\bm{z}) := \frac{\exp \, (z_i)}{\sum_{i = 1}^q \exp \, (z_i)}.
\label{Eq:Softmax}
\end{equation}
As a consequence, the order parameters in the compact state, $\phi^*$ and $\{ x_i^* \}_{i = 1}^q$, satisfy Eqs.~\eqref{Eq:OP_phi_Globule} and \eqref{Eq:OP_x_Globule}.

\subsubsection{Stretched state}

In the stretched state ($\phi^* = 0$ and $\eta^* > 0$), we need to minimize $f_\mathrm{S} (\eta, \{ x_i \})$ subject to $\sum_{i = 1}^q x_i = 1$.
Noticing that $\eta$ and $\{ x_i \}$ are uncoupled, we can easily obtain
\begin{equation}
\eta^* = \frac{\tilde{F}}{3}.
\label{Eq:OP_eta_Stretched}
\end{equation}
and
\begin{equation}
x_i^* = S_i \left( \tilde{\bm{h}} \right) \quad (1 \leq i \leq q),
\label{Eq:OP_x_Stretched}
\end{equation}
where $S_i \, (\bm{z})$ is the softmax function defined in Eq.~\eqref{Eq:Softmax}.

\subsubsection{Swollen state}
\label{Subsubsec:Swollen}

In the swollen state ($\phi^* = \eta^* = 0$), we only have to minimize $f_\mathrm{mark} (\{ x_i \})$ subject to $\sum_{i = 1}^q x_i = 1$.
The solution of the order parameters, $\{ x_i^* \}_{i = 1}^q$, is already obtained in Eq.~\eqref{Eq:OP_x_Stretched}.

We can derive a sufficient condition for the swollen state to be stabilized as explained in the following.
Assuming $F = 0$, we can see $\eta^* = 0$ since from Eq.~\eqref{Eq:PseudoFED_Stretched}, $f_\mathrm{S} (\eta, \{ x_i \})$ is a monotonously increasing function of $\eta$.
We thus show below a sufficient condition for $\phi^* = 0$ on the assumption that $F = 0$.
Focusing on the stationary condition of $\phi^*$ [Eq.~\eqref{Eq:OP_phi}], we can see that $\phi^*$ has to be zero if $\sum_{i, j} \tilde{J}_{ij} x_i x_j = {\bm{x}}^\mathrm{T} \, \tilde{\mathsf{J}} \, \bm{x}$ is smaller than $\tilde{v}$ for arbitrary values of $\{ x_i \}_{i = 1}^q$ satisfying $0 \leq x_i \leq 1$.
Introducing an orthogonal matrix $M$ which diagonalizes the symmetric matrix $\tilde{\mathsf{J}}$ as $\mathsf{M}^\mathrm{T} \, \tilde{\mathsf{J}} \, \mathsf{M} = \mathsf{\Lambda} := \mathrm{diag} (\lambda_1, \cdots, \lambda_q)$, writing the maximum eigenvalue of $\mathsf{\Lambda}$ as $\lambda_\mathrm{max}$, and defining $\bm{y} := \mathsf{M}^\mathrm{T} \bm{x}$, we obtain the following inequality,
\begin{equation}
\bm{x}^\mathrm{T} \, \tilde{\mathsf{J}} \, \bm{x} = \bm{y}^\mathrm{T} \, \mathsf{\Lambda} \, \bm{y} = \sum_{i = 1}^q \lambda_i {y_i}^2 \leq \lambda_\mathrm{max} \sum_{i = 1}^q {y_i}^2 = \lambda_\mathrm{max} \sum_{i = 1}^q {x_i}^2 \leq \lambda_\mathrm{max} \sum_{i = 1}^q x_i = \lambda_\mathrm{max}.
\end{equation}
The second inequality is due to $0 \leq x_i \leq 1$.
Therefore, if the maximum eigenvalue of $\tilde{\mathsf{J}}$, $\lambda_\mathrm{max}$, is smaller than $\tilde{v}$, the only solution of Eq.~\eqref{Eq:OP_phi} is $\phi^* = 0$.
In short, a sufficient condition for the swollen state to be realized is that the maximum eigenvalue of $\mathsf{J}$ is smaller than $v$ (i.e., interaction is effectively repulsive) and also $F = 0$ (i.e., no stretching force).

\subsection{Criteria for second-order transition}
\label{Subsec:2ndOrderTransPoint}

Let us consider the case with some external field ($\bm{h} \neq \bm{0}$) and no external force ($F = 0$), and explore under what conditions the second-order transition between a swollen state ($\phi^* = 0$) and a compact state ($\phi^* > 0$) can occur.

Recall first that in the swollen state, $\phi^* = 0$ and $x_i^* = S_i (\tilde{\bm{h}})$ (Sec.~\ref{Subsubsec:Swollen}).
Then, if we assume that a continuous transition occurs at $\tilde{J}_{ij} = \tilde{J}_{ij}^\mathrm{(c)}$, the system will smoothly change from the swollen state to the compact state in response to a certain small change in $\tilde{J}_{ij}$: $\phi^* = 0 + \Delta \phi$ and $x_i^* = S_i (\tilde{\bm{h}}) + \Delta x_i$ for $\tilde{J}_{ij} = \tilde{J}_{ij}^\mathrm{(c)} + \Delta \tilde{J}_{ij}$.
The pseudo free energy for $\tilde{J}_{ij} = \tilde{J}_{ij}^\mathrm{(c)} + \Delta \tilde{J}_{ij}$ is obtained from Eq.~ \eqref{Eq:PseudoFED_Globule} as
\begin{eqnarray}
\frac{{f}_\mathrm{C} (\phi, \{ x_i \})}{k_\mathrm{B} T} &=& \left\{ \tilde{v} - \sum_{i,j} \left( \tilde{J}_{ij}^\mathrm{(c)} + \Delta \tilde{J}_{ij} \right) \left[ S_i \left( \tilde{\bm{h}} \right) + \Delta x_i \right] \left[ S_j \left( \tilde{\bm{h}} \right) + \Delta x_j \right] \right\} \Delta \phi^2 + \Delta \phi^4 \nonumber \\
&& - \tilde{h}_i \left[ S_i \left( \tilde{\bm{h}} \right) + \Delta x_i \right] + \sum_{i = 1}^q \left[ S_i \left( \tilde{\bm{h}} \right) + \Delta x_i \right] \ln \left[ S_i \left( \tilde{\bm{h}} \right) + \Delta x_i \right].
\end{eqnarray}
Minimizing this function with respect to $\Delta \phi$, we obtain
\begin{equation}
\Delta \phi = \sqrt{\frac{1}{2} \left\{ \sum_{i,j} \left( \tilde{J}_{ij}^\mathrm{(c)} + \Delta \tilde{J}_{ij} \right) \left[ S_i \left( \tilde{\bm{h}} \right) + \Delta x_i \right] \left[ S_j \left( \tilde{\bm{h}} \right) + \Delta x_j \right] - \tilde{v} \right\}}.
\end{equation}
The assumption of a continuous transition leads to the continuity of the order parameters at $\tilde{J}_{ij} = \tilde{J}_{ij}^\mathrm{(c)}$: $\Delta \phi \to 0$ and $\Delta x_i \to 0$ for $\Delta \tilde{J}_{ij} \to 0$; therefore, we obtain the following expression of the possible second-order transition point:
\begin{equation}
\sum_{i, j} \tilde{J}_{ij}^\mathrm{(c)} S_i \left( \tilde{\bm{h}} \right) S_j \left( \tilde{\bm{h}} \right) = \tilde{v}.
\label{Eq:SecondOrderTransPoint}
\end{equation}
This means that within the modified polymer-Potts model [Eq.~\eqref{Eq:PolymerPottsLongRange}], any second-order conformation transition accompanying the changes in the order parameters should satisfy Eq.~\eqref{Eq:SecondOrderTransPoint}.
Conversely, if Eq.~\eqref{Eq:SecondOrderTransPoint} is not satisfied, the continuous swollen-compact phase transition cannot occur.
We will see in the following subsection~\ref{Subsec:SimpleCases} that a discontinuous transition can occur even if Eq.~\eqref{Eq:SecondOrderTransPoint} is not satisfied.

\subsection{Analysis of simple cases}
\label{Subsec:SimpleCases}

In this subsection, we provide some examples to explain how the equilibrium states and transition properties of the polymer-Potts model can change according to the histone-mark specificity of the interactions, $\{ J_{ij} \}$.
For simplicity, we assume $F = 0$ and $h_i = 0$.

\begin{figure}[t]
\includegraphics[scale=0.45]{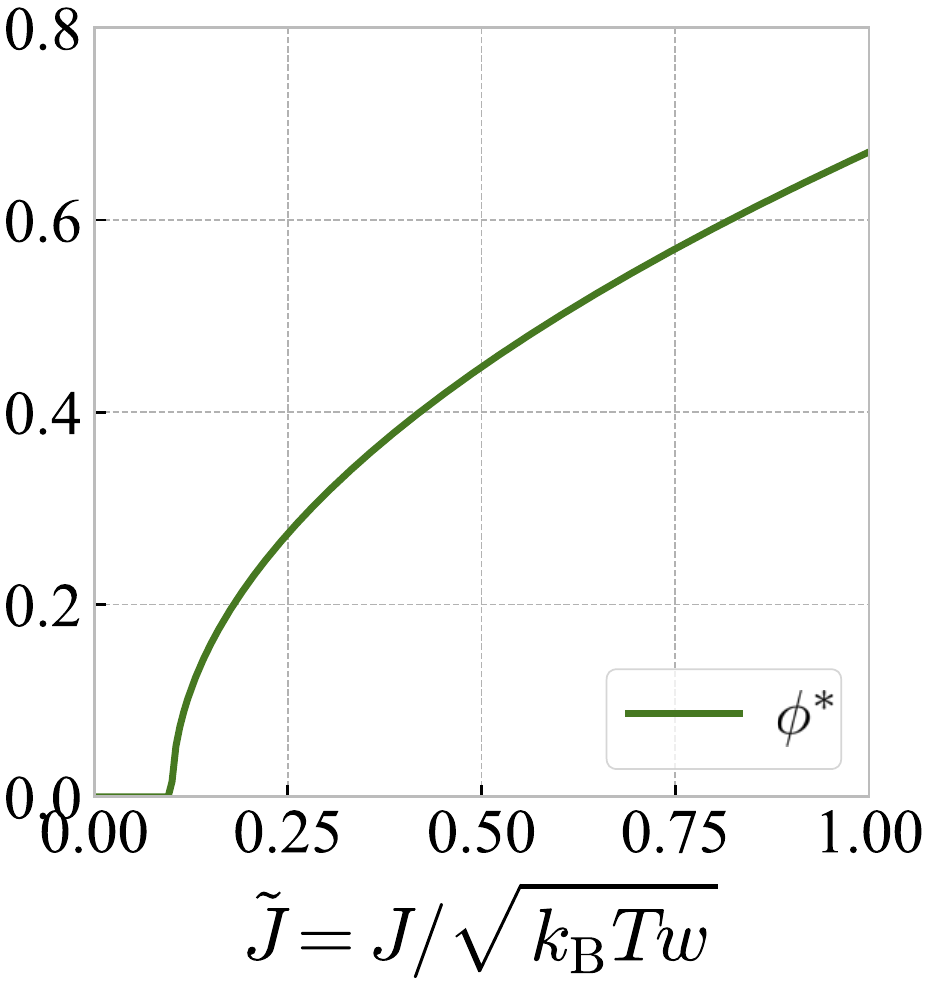}
\caption{Compact-state order parameter ($\phi^*$) as a function of the interaction strength ($\tilde{J}$) in the case of the mark-independent interaction and $\tilde{v} = 0.1$.}
\label{Fig:OPvsJ_Uniform_v01}
\end{figure}

\subsubsection{Mark-independent interaction}
\label{Subsubsec:NonSpecific}

Let us consider interactions which are not specific to histone marks ($J_{ij} = J$).
In this case, the formula of the free energy [Eq.~\eqref{Eq:FreeEnergyDensityFinal}] is reduced to $\lim_{N \to \infty} \overline{f} = \overline{f}_\mathrm{C} + \lim_{N \to \infty} f_0$ with the following pseudo free energy:
\begin{equation}
\frac{f_\mathrm{C} (\phi, \{ x_i \})}{k_\mathrm{B} T} = \left( \tilde{v} - \tilde{J} \right) \phi^2 + \phi^4 + \sum_{i = 1}^q x_i \ln x_i.
\end{equation}
The third term originated from the entropy of the marks can be minimized with $x_i = 1 / q$, so that we obtain the further reduced expression, $\lim_{N \to \infty} \overline{f} = \min_{\phi} f_\mathrm{C, U} (\phi) + \mathrm{const.}$, where the pseudo free energy for a uniform interaction, $f_\mathrm{C, U} (\phi)$, is defined as
\begin{equation}
f_\mathrm{C, U} (\phi) := \left( \tilde{v} - \tilde{J} \right) \phi^2 + \phi^4.
\end{equation}
Minimizing this function, we can see that there is a continuous transition as the interaction strength $\tilde{J}$ is varied through a critical point, $\tilde{J}^\mathrm{(c)} = \tilde{v}$, and obtain the stable equilibrium states as follows:
\begin{empheq}[left=\empheqlbrace]{align}
& \text{Continuous transition at } \tilde{J} = \tilde{J}^\mathrm{(c)} = \tilde{v}
\label{Eq:TransitionMarkIndep}
\\& \tilde{J} < \tilde{J}^\mathrm{(c)} \Rightarrow \text{Swollen-disordered state: } \phi^* = 0, \, x_i^* = \frac{1}{q} \ (1 \leq i \leq q) \\
& \tilde{J} > \tilde{J}^\mathrm{(c)} \Rightarrow \text{Compact-disordered state: } \phi^* = \sqrt{\frac{\tilde{J} - \tilde{J}^\mathrm{(c)}}{2}}, \, x_i^* = \frac{1}{q} \ (1 \leq i \leq q).
\end{empheq}
The obtained value of $\tilde{J}^\mathrm{(c)}$ is consistent with the discussion in the subsection~\ref{Subsec:2ndOrderTransPoint} [take $\bm{h} = \bm{0}$ and $\tilde{J}_{ij}^\mathrm{(c)} = \tilde{J}^\mathrm{(c)}$ in Eq.~\eqref{Eq:SecondOrderTransPoint}].
Figure~\ref{Fig:OPvsJ_Uniform_v01} shows the $\tilde{J}$ dependence of the order parameter $\phi^*$ for the case of $\tilde{v} = 0.1$.

Note that the mean-field model of the coil-globule transition introduced in \cite{DeGennes1975} is reduced to the same functional form as $f_\mathrm{C, U} (\phi)$ in the limit of $N \to \infty$ (see also Sec.~\ref{Sec:PolymerLR}).
Thus, with a uniform Potts-like interaction, the marks are always disordered ($x_i^* = 1 / q$), and the properties of the polymer-Potts model are reduced to those of a usual polymer model with two-body and three-body monomer-monomer interactions.

\begin{figure}[t]
\includegraphics[scale=0.45]{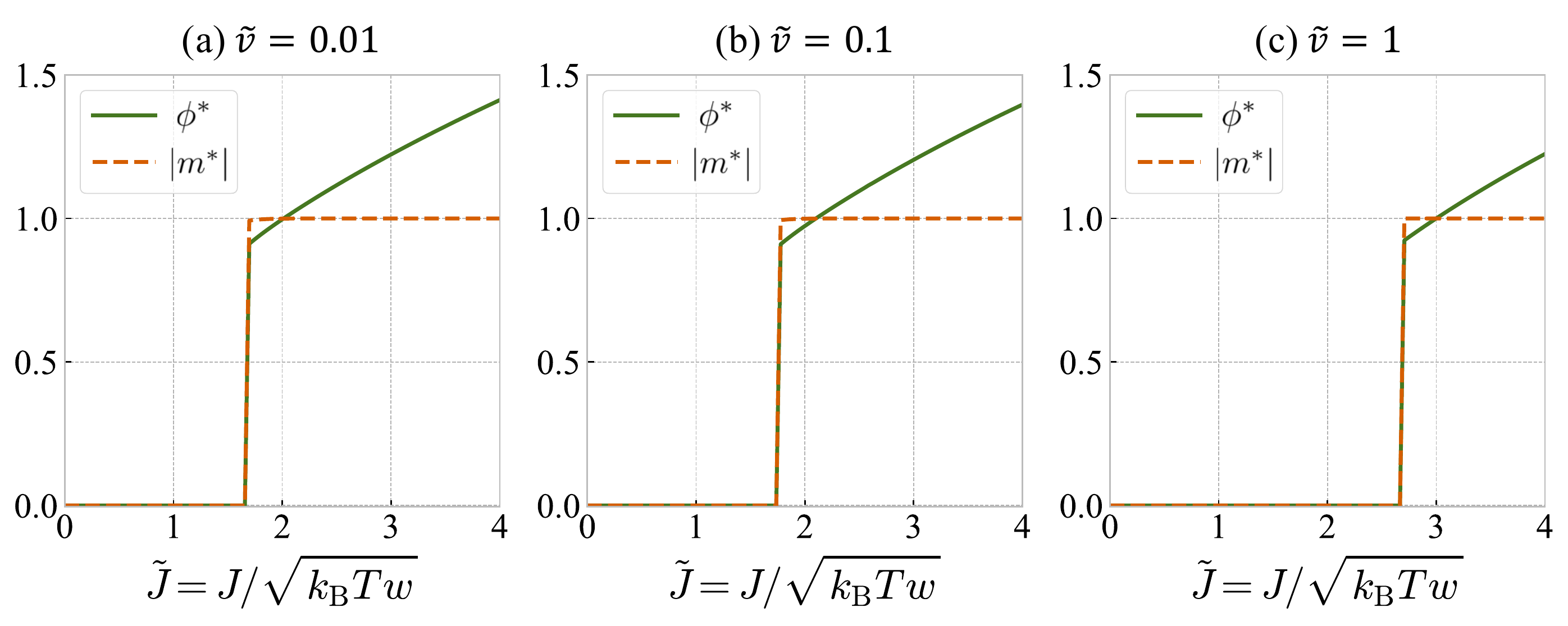}
\caption{Two kinds of order parameters ($\phi^*$ and $m^* = x_1^* - x_2^*$) as a function of the interaction strength ($\tilde{J}$) in the case of the Ising-type interaction.}
\label{Fig:OrderParams}
\end{figure}

\subsubsection{Mark-specific interaction: Ising-type}
\label{Subsubsec:Ising}

Here and in the main text, we consider one of the simplest histone mark-specific interaction, the Ising-type interaction: $J_{ij} =  J ( 2\delta_{ij} -1 )$ with $q = 2$.
According to the general discussion in the subsection~\ref{Subsec:2ndOrderTransPoint} [see Eq.~\eqref{Eq:SecondOrderTransPoint}], there is no second-order transition in this system since the possible second-order transition point cannot be achieved irrespective of the value of $\tilde{J} = J / \sqrt{k_\mathrm{B} T w}$.

It is natural to redefine the mark order parameter with the magnetization, $m := x_1 - x_2$.
The numerically obtained order parameters are shown in Fig.~\ref{Fig:OrderParams}.
In stark contrast to the non-mark-specific interaction case discussed in the subsection~\ref{Subsubsec:NonSpecific}, there is a first-order transition between the swollen-disordered state ($\phi^* = 0$, $m^* = 0$) and the compact-ordered state ($\phi^* > 0$ and $|m^*| > 0$) at a $\tilde{v}$-dependent transition point, $\tilde{J}^\mathrm{(c)}$.

Since the magnetization amplitude $|m^*|$ shows a jump from $0$ to $\simeq \! \! 1$, we can obtain an approximate formula of the first-order transition point $J^\mathrm{(c)}$ and that of the $\tilde{J}$ dependence of $\phi^*$ in the compact state, as explained in the following.
According to Eqs.~\eqref{Eq:FreeEnergyDensityFinal}, \eqref{Eq:FreeEnergyDensity_Compact}, and \eqref{Eq:PseudoFED_Globule}, the free energy of the swollen-disordered state is given as $\overline{f}_\mathrm{C} = f_\mathrm{C} (\phi = 0, x_1 = x_2 = 1 / 2) = -\ln 2$.
On the other hand, the free energy of the compact-ordered state is given as $\overline{f}_\mathrm{C} \simeq \min_\phi f_\mathrm{C} (\phi, x_1 = 1, x_2 = 0) \, [= \min_\phi f_\mathrm{C} (\phi, x_1 = 0, x_2 = 1)] = \min_\phi [(\tilde{v} - \tilde{J}) \phi^2 + \phi^4]$.
The last expression can be minimized with $\phi = \sqrt{(\tilde{J} - \tilde{v}) / 2}$, and finally we obtain in the compact-ordered state $\overline{f}_\mathrm{C} \simeq - (\tilde{J} - \tilde{v})^2 / 4$.
At the transition point ($\tilde{J} = \tilde{J}^\mathrm{(c)}$), the free energy of the swollen-disordered state and that of the compact-ordered state coincide with each other: $- \ln 2 = - (\tilde{J}^\mathrm{(c)} - \tilde{v})^2 / 4$, and we thus obtain $\tilde{J}^\mathrm{(c)} \simeq \tilde{v} + 2 \sqrt{\ln 2}$.

To summarize, the stable equilibrium state for the Ising-type interaction is given as follows:
\begin{empheq}[left=\empheqlbrace]{align}
& \text{Discontinuous transition at } \tilde{J} = \tilde{J}^\mathrm{(c)} \simeq \tilde{v} + 2 \sqrt{\ln 2}
\label{Eq:TransitionFirstOrder}
\\
& \tilde{J} < \tilde{J}^\mathrm{(c)} \Rightarrow \text{Swollen-disordered state: } \phi^* = 0, \ x_i^* = \frac{1}{2} \ (i = 1, 2) \\
& \tilde{J} > \tilde{J}^\mathrm{(c)} \Rightarrow \text{Compact-ordered state: } \phi^* \simeq \sqrt{\frac{\tilde{J} - \tilde{v}}{2}}, \ \max_i \{ x_i^* \} \simeq 1, \ \min_i \{ x_i^* \} \simeq 0.
\end{empheq}

Let us shortly consider the case with finite fields ($h_1 = h$ and $h_2 = -h$).
Under sufficiently high fields, the possible second-order transition point is given from Eq.~\eqref{Eq:SecondOrderTransPoint} as
\begin{equation}
\tilde{J}^\mathrm{(c)} (h) = \frac{\tilde{v}}{[\tanh (h / k_\mathrm{B} T)]^2}.
\label{Eq:SecondOrderField}
\end{equation}
As $h$ gets larger and the mark type becomes virtually fixed, $\tilde{J}^\mathrm{(c)} (h)$ gets close to $\tilde{v}$~\cite{Garel1999}, which is the second-order transition point of a homogeneous polymer with only mark-independent interactions [see Sec.~\ref{Subsubsec:NonSpecific} and Eq.~\eqref{Eq:TransitionMarkIndep}].
In addition, for low fields where the magnetization jump is expected around unity as in the zero-field case, we can estimate the first-order transition point as in Eq.~\eqref{Eq:TransitionFirstOrder}:
\begin{equation}
\tilde{J}^\mathrm{(c)} (h) \simeq \tilde{v}+ 2 \sqrt{\ln 2 - \frac{h}{k_\mathrm{B} T}}
\label{Eq:FirsrOrderField}
\end{equation}
These expressions [Eqs.~\eqref{Eq:SecondOrderField} and \eqref{Eq:FirsrOrderField}] are plotted in Fig.~3(b) in the main text and compared with the numerical results.

Lastly, we consider external-force effects.
In this case, the first-order transition occurs as a function of $F$ and $J$.
Since the magnetization jump is almost unity, we can estimate the first-order transition point in a similar way to Eq.~\eqref{Eq:TransitionFirstOrder}:
\begin{equation}
F^\mathrm{(c)} (\tilde{J}) b / k_\mathrm{B} T \simeq \sqrt{\frac{3 (\tilde{J} - \tilde{v})^2}{2} - 6 \ln 2},
\end{equation}
which is also compared with the numerical results [Fig.~3(f) in the main text].

\begin{figure}[t]
\begin{tabular}{c}\begin{minipage}{1\hsize}
\includegraphics[scale=0.45]{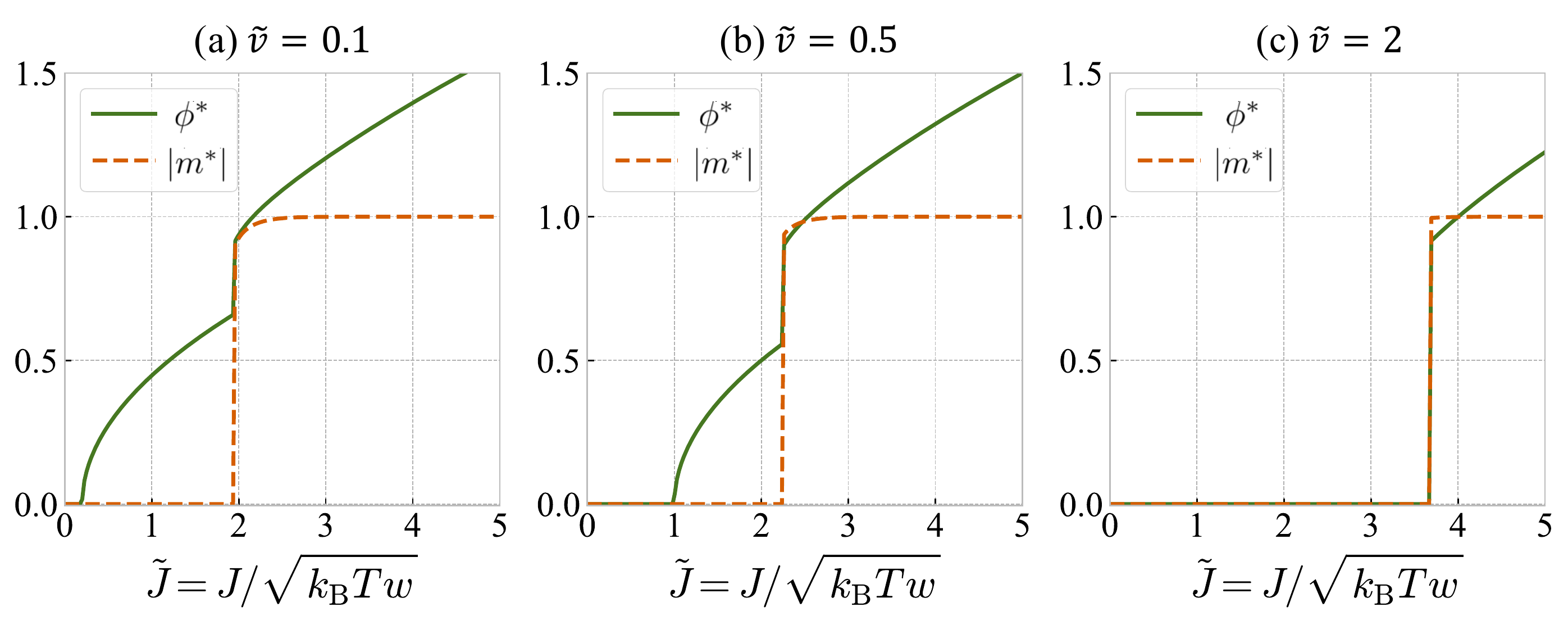}
\caption{Order parameters ($\phi^*$ and $m^* = x_1^* - x_2^*$) as a function of the interaction strength ($\tilde{J}$) in the case of the $2$-state ($q = 2$) Potts-type interaction.}
\label{Fig:OrderParams_2Potts}
\end{minipage} \vspace{20pt} \\
\begin{minipage}{1\hsize}
\includegraphics[scale=0.45]{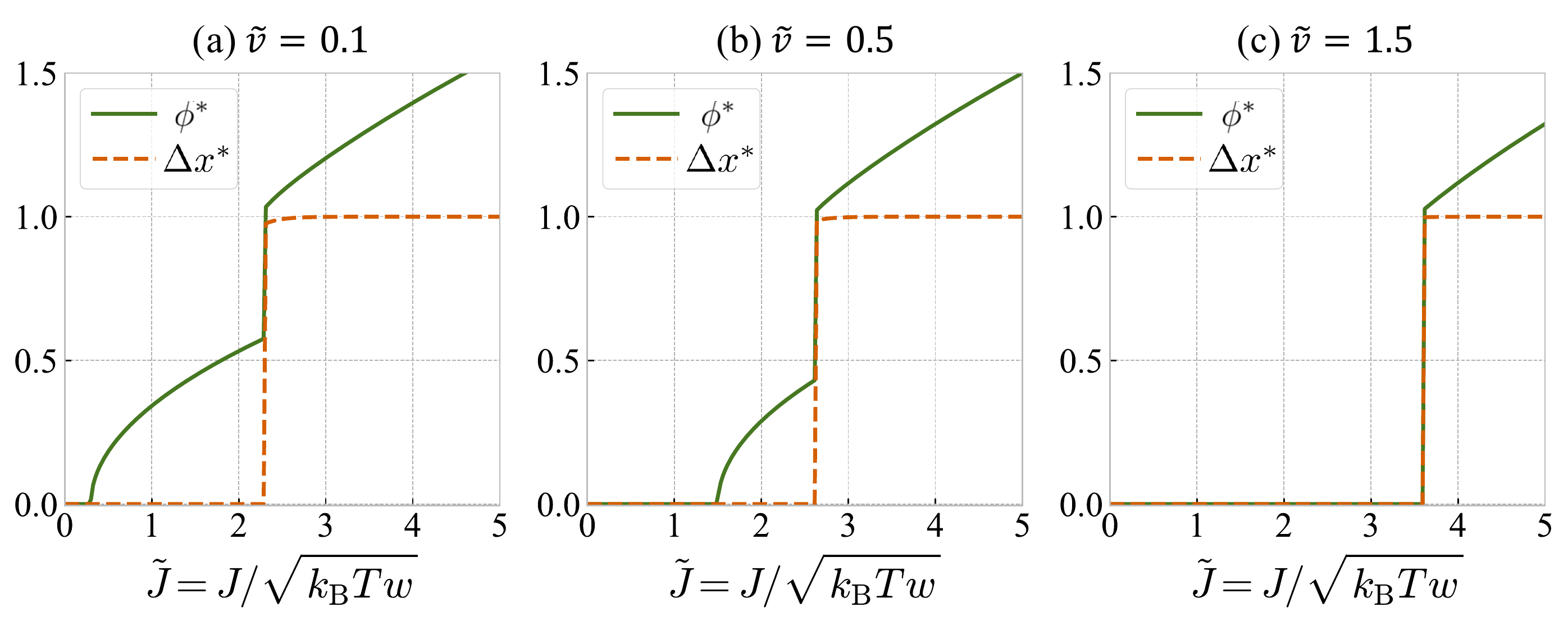}
\caption{Order parameters ($\phi^*$ and $\Delta x^* = \max_i \{ x_i^* \} - \min_i \{ x_i^* \}$) as a function of the interaction strength ($\tilde{J}$) in the case of the $3$-state ($q = 3$) Potts-type interaction.}
\label{Fig:OrderParams_3Potts}
\end{minipage} \vspace{20pt} \\
\begin{minipage}{1\hsize}
\includegraphics[scale=0.5]{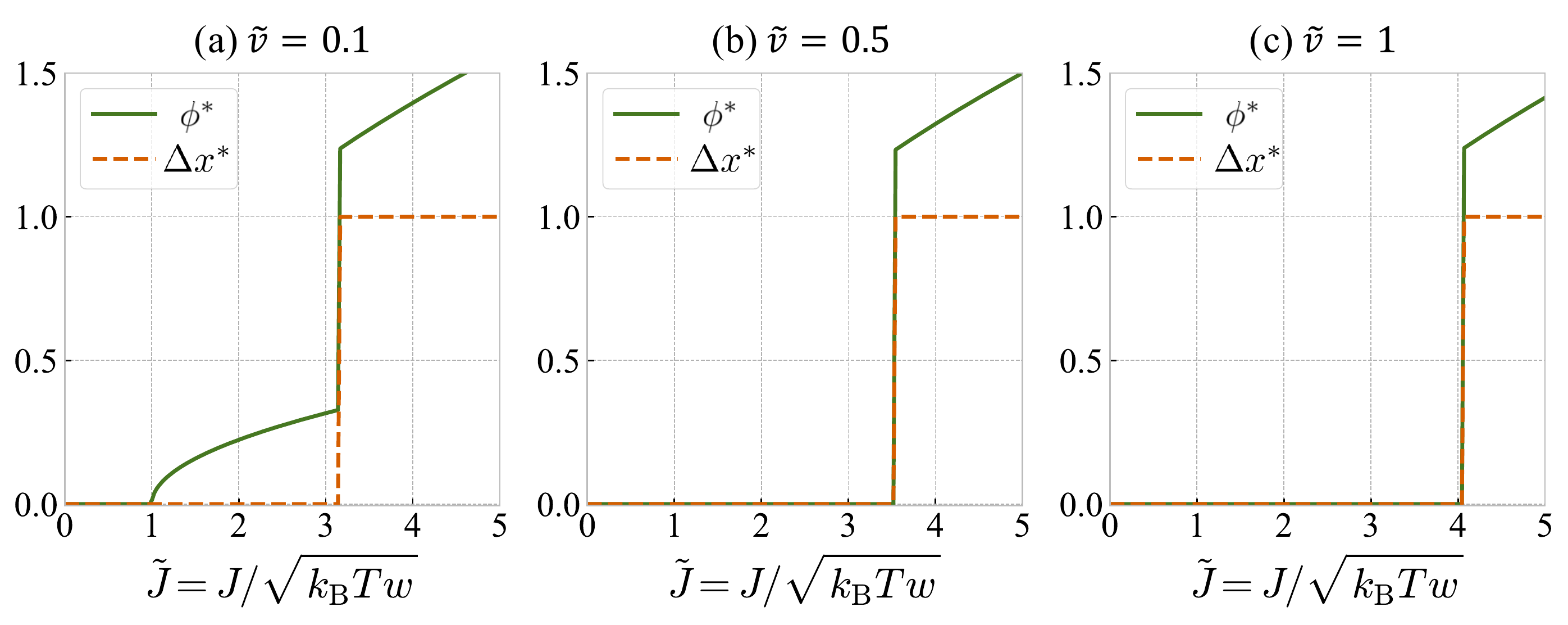}
\caption{Order parameters ($\phi^*$ and $\Delta x^* = \max_i \{ x_i^* \} - \min_i \{ x_i^* \}$) as a function of the interaction strength ($\tilde{J}$) in the case of the $10$-state ($q = 10$) Potts-type interaction.}
\label{Fig:OrderParams_10Potts}
\end{minipage}
\end{tabular}
\end{figure}

\begin{figure}[t]
\includegraphics[scale=0.45]{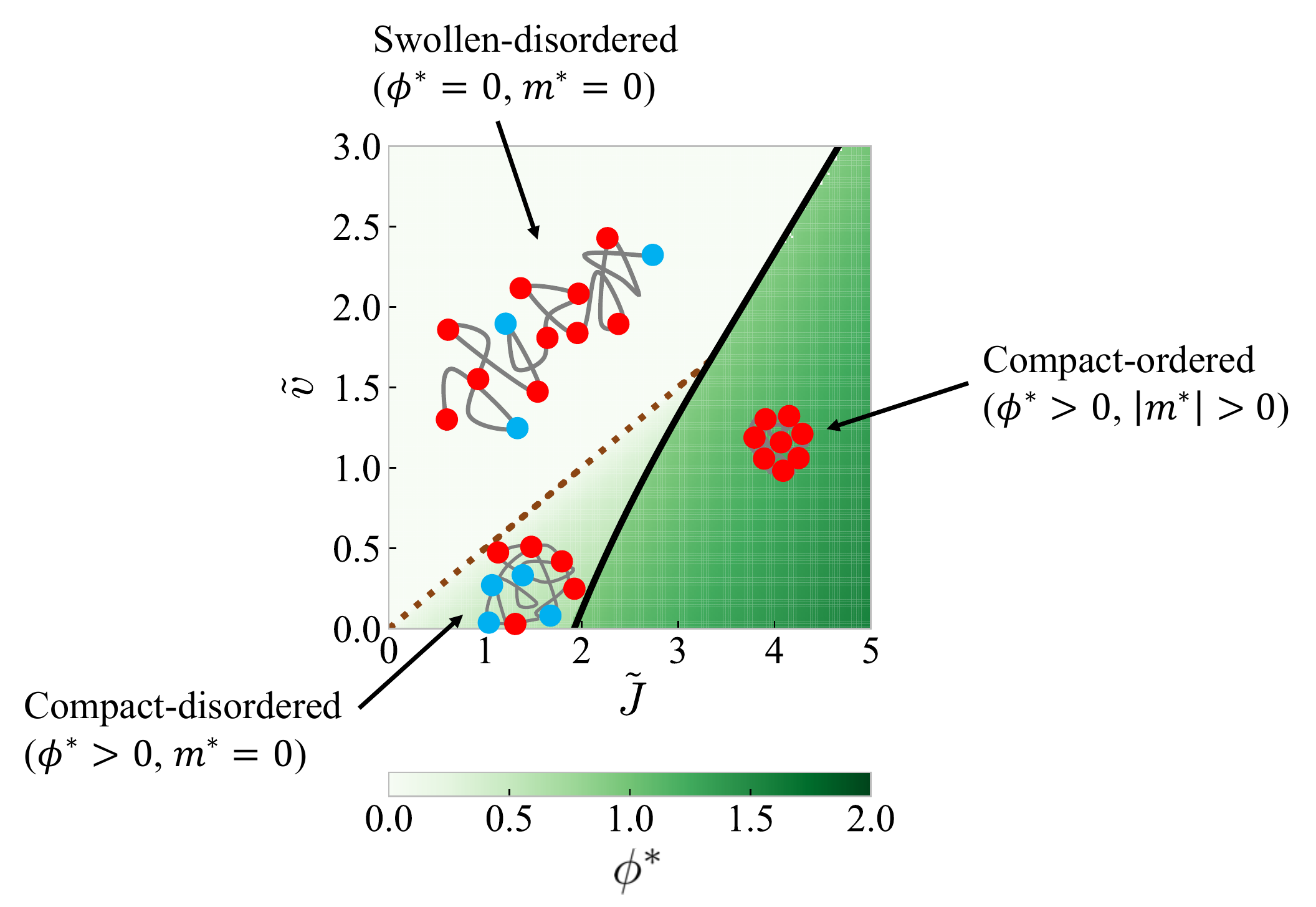}
\caption{Phase diagram in the $J$-$v$ plane for the case of 2-state ($q = 2$) Potts-type interaction.
The compact-state order parameter ($\phi^*$) is plotted as the color depth.
The continuous transition line [Eq.~\eqref{Eq:Jc_SD_CD}] between the swollen-disordered and the compact-disordered states is shown with a brown dotted line.
Also, the approximate discontinuous transition line between the swollen-disordered and compact-ordered states [Eq.~\eqref{Eq:Jc_SD_CO}] as well as that between the compact-disordered and compact-ordered states [Eq.~\eqref{Eq:Jc_CD_CO}] is shown with a black solid line.
The approximate cross point [Eqs.~\eqref{Eq:v_cross} and \eqref{Eq:J_cross}] is given as $\tilde{J}^\mathrm{(cross)} \simeq 3.33$ and $\tilde{v}^\mathrm{(cross)} \simeq 1.67$.}
\label{Fig:PhaseDiagram_2statePotts}
\end{figure}

\subsubsection{Mark-specific interaction: $q$-state Potts-type interaction}

As another example of a mark-specific interaction, let us think of the $q$-state Potts-type interaction: $J_{ij} = J \delta_{ij}$ with $q$ $(\geq 2)$ kinds of marks.
In contrast to the Ising-type interaction case, according to Eq.~\eqref{Eq:SecondOrderTransPoint}, there is a possible second-order transition at 
\begin{equation}
\tilde{J}^\mathrm{(c)}_\mathrm{2nd} = q \tilde{v} \quad \text{(from swollen-disordered to compact-disordered)}
\label{Eq:Jc_SD_CD}
\end{equation}

In Fig.~\ref{Fig:OrderParams_2Potts}, the $\tilde{J}$ dependence of the order parameters, $\phi^*$ and $m^* := x^*_1 - x^*_2$, is shown for the case of $q = 2$.
For small values of $\tilde{v}$ [Figs.~\ref{Fig:OrderParams_2Potts}(a) and (b)], two-step transitions occur: the compact-disordered state ($\phi^* > 0$, $m^* = 0$) starts to appear at $\tilde{J} = \tilde{J}^\mathrm{(c)}_\mathrm{2nd}$ through a second-order transition, and then the compact-ordered state ($\phi^* > 0$ and $|m^*| > 0$) emerges at $\tilde{J} = \tilde{J}^\mathrm{(c)}$ through a first-order transition.
For large values of $\tilde{v}$ [Fig.~\ref{Fig:OrderParams_2Potts}(c)], on the other hand, a single-step transition occurs: the compact-ordered state emerges at $\tilde{J} = \tilde{J}^\mathrm{(c)}$ through a first-order transition.
Defining the magnetization as $\Delta x^* := \max_i \{ x_i^* \} - \min_i \{ x_i^* \}$, we can see that such qualitative features of the order parameters seem to be invariant even when the number of different marks ($q$) is larger than $2$ (see Figs.~\ref{Fig:OrderParams_3Potts} and \ref{Fig:OrderParams_10Potts} for the case of $q = 3$ and $q = 10$, respectively).

For the case of large $\tilde{v}$, noticing that the magnetization size ($\Delta x^*$) shows a jump from $0$ to $\simeq \! \! 1$  we can estimate the first-order transition point ($\tilde{J}^\mathrm{(c)}$), just as in the case of the Ising-type interaction (see the subsection~\ref{Subsubsec:Ising}).
Comparing the free energy in the swollen-disordered state and that of the compact-ordered state, we can obtain
\begin{equation}
\tilde{J}^\mathrm{(c)} \simeq \tilde{v} + 2 \sqrt{\ln q} \quad \text{(from swollen-disordered to compact-ordered)}.
\label{Eq:Jc_SD_CO}
\end{equation}
On the other hand, for the case of small $\tilde{v}$, a similar comparison of the free-energy difference between the compact-disordered state and the compact-ordered state leads to 
$-(\tilde{J} / q - \tilde{v})^2 / 4 - \ln q = -(\tilde{J} - \tilde{v})^2 / 4$, and we then obtain
\begin{equation}
\tilde{J}^\mathrm{(c)} \simeq \frac{q \tilde{v}}{q + 1} + \frac{q}{q + 1} \sqrt{\tilde{v}^2 + 4 \, \frac{q + 1}{q - 1} \ln q} \quad \text{(from compact-disordered to compact-ordered)}.
\label{Eq:Jc_CD_CO}
\end{equation}
The approximate cross point ($\tilde{v}^\mathrm{(cross)}$, $\tilde{J}^\mathrm{(cross)}$) of the first-order and second-order transition lines can be estimated with Eqs.~\eqref{Eq:Jc_SD_CD} and \eqref{Eq:Jc_SD_CO} [or Eqs.~\eqref{Eq:Jc_SD_CD} and \eqref{Eq:Jc_CD_CO}] as
\begin{empheq}[left=\empheqlbrace]{align}
& \tilde{v}^\mathrm{(cross)} \simeq \frac{2 \sqrt{\ln q}}{q - 1}
\label{Eq:v_cross} \\
& \tilde{J}^\mathrm{(cross)} = q \tilde{v}^\mathrm{(cross)} \simeq \frac{2 q \sqrt{\ln q}}{q - 1}.
\label{Eq:J_cross}
\end{empheq}

The numerically acquired phase diagram in the $J$-$v$ plane for the case of $q = 2$ is shown in Fig.~\ref{Fig:PhaseDiagram_2statePotts} with the analytic expressions of the transition lines [Eqs.~\eqref{Eq:Jc_SD_CD}, \eqref{Eq:Jc_SD_CO}, and \eqref{Eq:Jc_CD_CO}].

\section{Polymer model with long-range interactions}
\label{Sec:PolymerLR}

In this section, we consider a polymer model with long-range interactions and show that this model is essentially equivalent to the mean-field model of the coil-globule transition introduced in \cite{DeGennes1975}.

Let us consider a polymer model with long-range interactions:
\begin{equation}
H^\mathrm{LR}_\mathrm{polymer} = H_\mathrm{GC} (\{ {\bm R}_n \}) + H_\mathrm{VE}^\mathrm{LR} (\bm{R}_{N0}),
\end{equation}
where the Gaussian-chain term, $H_\mathrm{GC} (\{ {\bm R}_n \})$, is defined in Eq.~\eqref{Eq:GaussianChain}, and the long-range volume exclusion interaction term, $H_\mathrm{VE}^\mathrm{LR} (\bm{R}_{N0})$, is defined in Eq.~\eqref{Eq:VolumeExclusionLR}.
In this section, we assume that $v$ can be positive or negative, while $w$ is positive.

The equilibrium partition function, $Z_\mathrm{polymer}$, is defined as
\begin{equation}
Z_\mathrm{polymer} := \left( \prod_{n = 1}^N \int \mathrm{d} \bm{R}_n \right) \exp \left( -\frac{H^\mathrm{LR}_\mathrm{polymer}}{k_\mathrm{B} T} \right).
\end{equation}
Introducing the end-to-end vector $\bm{R}$ with an identity, $\int \mathrm{d} \bm{R} \, \delta (\bm{R} - \bm{R}_{N0}) = 1$, we can rewrite $Z_\mathrm{polymer}$ in a similar way to Eq.~\eqref{Eq:CalculationIdentity} as
\begin{eqnarray}
Z_\mathrm{polymer} &=& \int \mathrm{d} \bm{R} \exp \left( - \frac{v'}{k_\mathrm{B} T} \frac{N^2}{R^3} - \frac{w'}{k_\mathrm{B} T} \frac{N^3}{R^6} \right) \nonumber \\
&& \times \left( \prod_{n = 1}^N \int \mathrm{d} \bm{R}_n \right) \delta (\bm{R} - \bm{R}_{N0}) \exp \left[ - \frac{3}{2 b^2} \sum_{n = 1}^N (\bm{R}_n - \bm{R}_{n - 1})^2 \right],
\end{eqnarray}
where $v' = 3 v / 4 \pi$ and $w' = (3 / 4 \pi)^2 w$ as in Sec.~\ref{Sec:PolymerPotts}.
Through the same calculation process as Eq.~\eqref{Eq:CalculationGC}, we obtain
\begin{equation}
Z_\mathrm{polymer} = \frac{1}{N^{3/2}} \left( \frac{2 \pi b^2}{3} \right)^{3(N - 1) / 2} \int \mathrm{d} \bm{R} \exp \left( - \frac{3 R^2}{2 N b^2} - \frac{v'}{k_\mathrm{B} T} \frac{N^2}{R^3} - \frac{w'}{k_\mathrm{B} T} \frac{N^3}{R^6} \right).
\end{equation}
Performing the solid-angle integration of $\bm{R}$ and exponentiating all the factors lead to the following form:
\begin{equation}
Z_\mathrm{polymer} = \int_0^\infty \mathrm{d} R \exp \left[ \frac{3 (N - 1)}{2} \ln \frac{2 \pi b^2}{3} - \frac{3}{2} \ln N + \ln (4 \pi) + 2 \ln R - \frac{3 R^2}{2 N b^2} - \frac{v'}{k_\mathrm{B} T} \frac{N^2}{R^3} - \frac{w'}{k_\mathrm{B} T} \frac{N^3}{R^6} \right].
\end{equation}

Following \cite{DeGennes1975}, we define the expansion factor, $\alpha := R / \sqrt{N} b$, which leads to
\begin{equation}
Z_\mathrm{polymer} = \int_0^\infty \mathrm{d} \alpha \exp \left[ - \frac{F_\mathrm{eff} (\alpha)}{k_\mathrm{B} T} + \mathrm{const.} \right],
\end{equation}
where $F_\mathrm{eff} (\alpha)$ is an effective free energy  defined as
\begin{equation}
\frac{F_\mathrm{eff} (\alpha)}{k_\mathrm{B} T} := \frac{3}{2} \alpha^2 - 2 \ln \alpha + \frac{x}{\alpha^3} + \frac{y}{2 \alpha^6}
\label{Eq:EffectiveFreeEnergy}
\end{equation}
with $x := v' \sqrt{N} / k_\mathrm{B} T b^3$ and $y := 2 w' / k_\mathrm{B} T b^6$.
The effective free energy, $F_\mathrm{eff} (\alpha)$, has the same functional form as the mean-field free energy, Eq.~(1) in \cite{DeGennes1975}, apart from the numerical coefficient of the second term [$-3 \ln \alpha$ in \cite{DeGennes1975} instead of $-2 \ln \alpha$ in Eq.~\eqref{Eq:EffectiveFreeEnergy}].
This numerical difference of the coefficient only produces quantitative difference in the coil-globule transition properties for finite $N$, and moreover, the term proportional to $\ln \alpha$ becomes irrelevant in the limit of $N \to \infty$ since $\ln \alpha = o (N)$ (see also the discussion in the subsection~\ref{Subsubsec:NonSpecific}).
Therefore, the polymer model with long-range two-body and three-body interactions is essentially equivalent to the mean-field free energy introduced in \cite{DeGennes1975}.

\section{Relation to microscopic models}
\
In this section, we employ the virial expansion method to obtain a relation between the pseudo free energy of the polymer-Potts model and a microscopic Hamiltonian describing a polymer with mark-dependent interactions.

\subsection{Gas model}

Let us consider the microscopic two-body interaction term in the Hamiltonian of a virtual gas of $q$ types of monomers:
\begin{equation}
H_\mathrm{int} = \sum_{1 \leq n < m \leq N} U (\bm{r}_n - \bm{r}_m; s_n, s_m)
\label{Eq:MicroInteraction}
\end{equation}
Here, $\bm{r}_n$ and $s_n$ are the position and mark of the $n$-th monomer, respectively, and $U (\bm{r} - \bm{r}'; s, s')$ is the interaction energy between one monomer with the mark $s$ placed at $\bm{r}$ and another monomer with the mark $s'$ placed at $\bm{r}'$.
A basic inversion symmetry, $U (\bm{r}_n - \bm{r}_m; s_n, s_m) = U (\bm{r}_m - \bm{r}_n; s_n, s_m) = U (\bm{r}_m - \bm{r}_n; s_m, s_n)$ is assumed.
In the following, $U (\bm{r}_n - \bm{r}_m; s_n, s_m)$ is also expressed as $U_{nm}$ for simplicity.

\subsection{Virial expansion}

Extending the standard procedure of the virial expansion~\cite{Mayer1977} to the multiple-mark case, we can formally expand the interaction part of the partition function $Q := V^{-N} \int \prod_{n=1}^N \mathrm{d}^3 \bm{r}_n \exp (- H_\mathrm{int} / k_\mathrm{B} T)$ as
\begin{eqnarray}
Q &=& \frac{1}{V^N} \int \left( \prod_{n = 1}^N \mathrm{d}^3 \bm{r}_n \right) \exp \left( - \frac{1}{k_\mathrm{B} T} \sum_{1 \leq n < m \leq N} U_{nm} \right) \nonumber \\
&=& \exp \left( \frac{1}{V^2} \sum_{1 \leq n < m \leq N} \int \mathrm{d}^3 \bm{r}_n \mathrm{d}^3 \bm{r}_m f_{nm} + \frac{1}{V^3} \sum_{1 \leq n < m < k \leq N} \int \mathrm{d}^3 \bm{r}_n \mathrm{d}^3 \bm{r}_m \mathrm{d}^3 \bm{r}_k f_{nm} f_{nk} f_{mk} + \cdots \right),
\end{eqnarray}
where
\begin{equation}
f_{nm} = f \left( \bm{r}_n - \bm{r}_m; s_n, s_m \right) = \exp \left( -\frac{U_{nm}}{k_\mathrm{B} T} \right) - 1
\end{equation}
is the Mayer f-function.

Using an abbreviation, $U^{(ij)}_{12} := U (\bm{r}_1 - \bm{r}_2; i, j)$ and correspondingly $f^{(ij)}_{12} := f (\bm{r}_1 - \bm{r}_2; i, j)$, we can obtain the second-order virial coefficient as
\begin{equation}
\frac{1}{V^2} \sum_{1 \leq n < m \leq N} \int \mathrm{d}^3 \bm{r}_n \mathrm{d}^3 \bm{r}_m f_{nm} = \sum_{1 \leq i \leq q} \frac{{N_i}^2}{2V^2} \int \mathrm{d}^3 \bm{r}_1 \mathrm{d}^3 \bm{r}_2 f^{(ii)}_{12} + \sum_{1 \leq i \neq j \leq q} \frac{N_i N_j}{2 V^2} \int \mathrm{d}^3 \bm{r}_1 \mathrm{d}^3 \bm{r}_2 f^{(ij)}_{12} + O (N^0).
\label{Eq:SecondVirial}
\end{equation}
Using similar expressions, we can also obtain the third-order virial coefficient as
\begin{eqnarray}
\frac{1}{V^3} \sum_{1 \leq n < m < k \leq N} \int \mathrm{d}^3 \bm{r}_n \mathrm{d}^3 \bm{r}_m \mathrm{d}^3 \bm{r}_k f_{nm} f_{nk} f_{mk} &=& \sum_{1 \leq i \leq q} \frac{{N_i}^3}{6 V^3} \int \mathrm{d}^3 \bm{r}_1 \mathrm{d}^3 \bm{r}_2 \mathrm{d}^3 \bm{r}_3 f^{(ii)}_{12} f^{(ii)}_{13} f^{(ii)}_{23} + \sum_{1 \leq i \neq j \leq q} \frac{{N_i}^2 N_j}{2 V^3} \int \mathrm{d}^3 \bm{r}_1 \mathrm{d}^3 \bm{r}_2 \mathrm{d}^3 \bm{r}_3 f^{(ii)}_{12} f^{(ij)}_{13} f^{(ij)}_{23} \nonumber \\
&& + \sum_{1 \leq i \neq j \neq k \neq i \leq q} \frac{N_i N_j N_k}{6 V^3} \int \mathrm{d}^3 \bm{r}_1 \mathrm{d}^3 \bm{r}_2 \mathrm{d}^3 \bm{r}_3 f^{(ij)}_{12} f^{(ik)}_{13} f^{(jk)}_{23} + O (N^0).
\label{Eq:ThirdVirial}
\end{eqnarray}
Higher-order virial coefficients can be obtained in a similar manner.

\subsection{Relation to polymer-Potts model with Ising-type interaction}
\label{Subsec:RelationVirial}

Here, let us consider the Ising-type case where $q = 2$ and $U(\bm{r}; 1, 1) = U(\bm{r}; 2, 2)$ (and thus $f_{12}^{(11)} = f_{12}^{(22)}$).
Taking $v$ and $J$ as
\begin{equation}
v = \frac{k_\mathrm{B} T}{V} \left( - \frac{1}{4} \int \mathrm{d}^3 \bm{r}_1 \mathrm{d}^3 \bm{r}_2 f^{(11)}_{12} - \frac{1}{4} \int \mathrm{d}^3 \bm{r}_1 \mathrm{d}^3 \bm{r}_2 f^{(12)}_{12} \right)
\label{Eq:Formula_v}
\end{equation}
and
\begin{equation}
J = \frac{k_\mathrm{B} T}{V} \left( \frac{1}{4} \int \mathrm{d}^3 \bm{r}_1 \mathrm{d}^3 \bm{r}_2 f^{(11)}_{12} - \frac{1}{4} \int \mathrm{d}^3 \bm{r}_1 \mathrm{d}^3 \bm{r}_2 f^{(12)}_{12} \right),
\label{Eq:Formula_J}
\end{equation}
we can readily obtain from Eq.~\eqref{Eq:SecondVirial}
\begin{equation}
\frac{1}{V^2} \sum_{1 \leq n < m \leq N} \int \mathrm{d}^3 \bm{r}_n \mathrm{d}^3 \bm{r}_m f_{nm} = -\frac{V}{k_\mathrm{B} T} \left( v \rho^2 - J m^2 \rho^2 \right) + O(N^0),
\label{Eq:2ndVirial_2state}
\end{equation}
where the density $\rho = N / V$ and the magnetization $m = (N_1 - N_2) / N$ are introduced.
The form of Eq.~\eqref{Eq:2ndVirial_2state} corresponds to the two-body interaction terms in the polymer-Potts pseudo free energy [Eq.~(1) in the main text] if we take $V \sim R^3$ with the polymer length $R$.

In a similar way, taking $w$ and $w'$ as
\begin{equation}
w = \frac{k_\mathrm{B} T}{V} \left( - \frac{1}{24} \int \mathrm{d}^3 \bm{r}_1 \mathrm{d}^3 \bm{r}_2 \mathrm{d}^3 \bm{r}_3 f^{(11)}_{12} f^{(11)}_{13} f^{(11)}_{23} - \frac{1}{8} \int \mathrm{d}^3 \bm{r}_1 \mathrm{d}^3 \bm{r}_2 \mathrm{d}^3 \bm{r}_3 f^{(11)}_{12} f^{(12)}_{13} f^{(12)}_{23} \right)
\label{Eq:Formula_w}
\end{equation}
and
\begin{equation}
w' = \frac{k_\mathrm{B} T}{V} \left( - \frac{1}{8} \int \mathrm{d}^3 \bm{r}_1 \mathrm{d}^3 \bm{r}_2 \mathrm{d}^3 \bm{r}_3 f^{(11)}_{12} f^{(11)}_{13} f^{(11)}_{23} + \frac{1}{8} \int \mathrm{d}^3 \bm{r}_1 \mathrm{d}^3 \bm{r}_2 \mathrm{d}^3 \bm{r}_3 f^{(11)}_{12} f^{(12)}_{13} f^{(12)}_{23} \right),
\label{Eq:Formula_wprime}
\end{equation}
we can obtain from Eq.~\eqref{Eq:ThirdVirial}
\begin{equation}
\frac{1}{V^3} \sum_{1 \leq n < m < k \leq N} \int \mathrm{d}^3 \bm{r}_n \mathrm{d}^3 \bm{r}_m \mathrm{d}^3 \bm{r}_k f_{nm} f_{nk} f_{mk} = - \frac{V}{k_\mathrm{B} T} \left( w \rho^3 + w' m^2 \rho^3 \right) + O(N^0).
\label{Eq:3rdVirial_2state}
\end{equation}
The first term of Eq.~\eqref{Eq:3rdVirial_2state} corresponds to the three-body interaction term in the polymer-Potts pseudo free energy [Eq.~(1) in the main text].

To sum up, the interaction part of the partition function can be represented as
\begin{equation}
Q = \exp \left[ -\frac{V}{k_\mathrm{B} T} \left( v \rho^2 - J m^2 \rho^2 + w \rho^3 + w' m^2 \rho^3 + O (\rho^4) \right) + O(N^0) \right].
\end{equation}
Thus, replacing $V = \mathrm{const.} \times R^3$ and adding the entropic contribution, we obtain the mean-field pseudo free energy (per monomer) of the polymer-Potts model in $N \to \infty$ limit:
\begin{equation}
\frac{f (\phi, m)}{k_B T} = \left( \tilde{v} - \tilde{J} m^2 \right) \phi^2 + \left( 1 + \tilde{w}' m^2 \right) \phi^4 + \frac{1}{2} \left[ \left( 1 + m \right) \ln \left( 1 + m \right) + \left( 1 - m \right) \ln \left( 1 - m \right) \right] + O \left( \phi^6 \right).
\label{Eq:PseudoFreeEnergy_Virial}
\end{equation}
Here, the dimensionless globular order parameter $\phi = (w / k_\mathrm{B} T)^{1/4} \sqrt{\rho}$ and the dimensionless coupling constants $\tilde{v} = v / \sqrt{k_\mathrm{B} T w}$, $\tilde{J} = J / \sqrt{k_\mathrm{B} T w}$, and $\tilde{w}' = w' / w$ are used as in the main text.
Note that the term proportional to $m^2 \phi^4$ does not appear in the minimal model considered in the main text.
Neglecting the higher-order terms $O(\phi^6)$ (although, strictly speacking, this operation is not justified when a discontinuous coil-globule transition occurs), we can discuss based on Eq.~\eqref{Eq:PseudoFreeEnergy_Virial} the phase transition nature of the polymer-Potts model with general interactions satisfying the Ising symmetry [$U (\bm{r}; 1, 1) = U (\bm{r}; 2, 2)$].

\begin{figure}[t]
\includegraphics[scale=0.5]{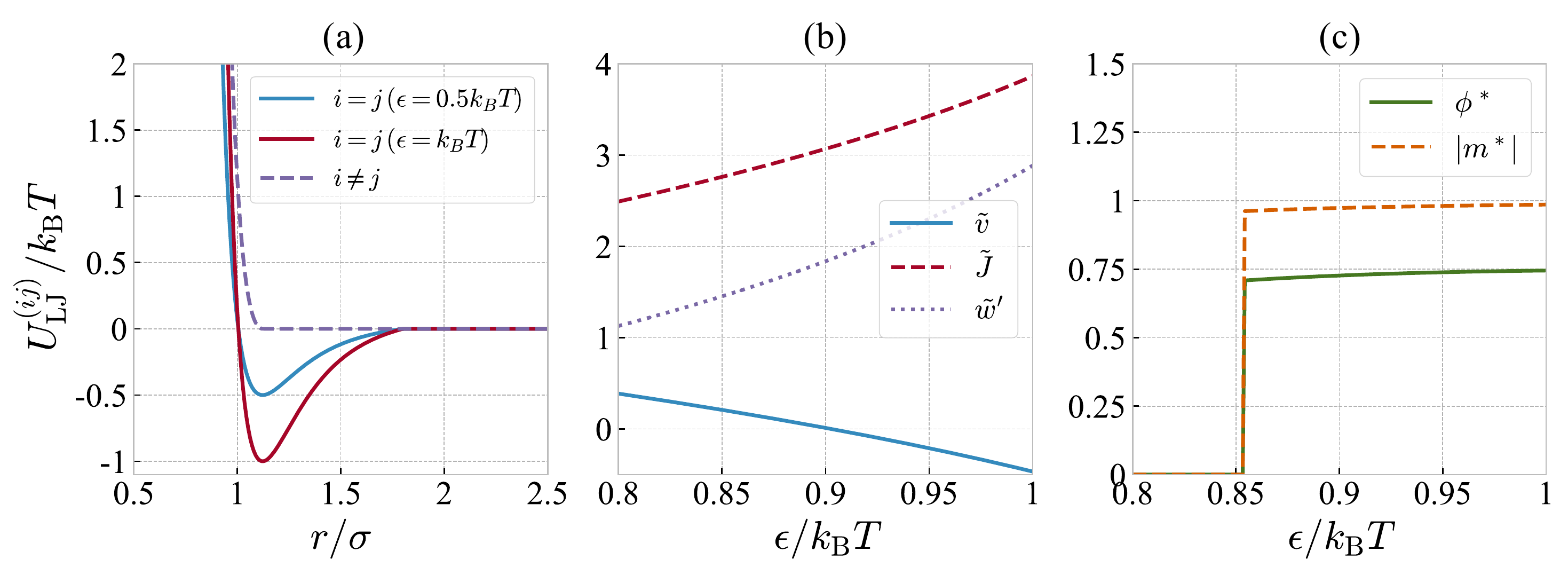}
\caption{(a)~Truncated Lennard-Jones-type interaction potential [Eq.~\eqref{Eq:InteractionLJ}].
(b)~Dimensionless coupling constants [$\tilde{v}$ (blue solid line), $\tilde{J}$ (red dashed line), and $\tilde{w}'$ (purple dotted line)] and (c)~equilibrium order parameters [$\phi^*$ (green solid line) and $m^*$ (orange dashed line)] as a function of the interaction strength $\epsilon / k_\mathrm{B} T$ for the case of the Lennard-Jones-type interaction [Eq.~\eqref{Eq:InteractionLJ}].}
\label{Fig:Virial}
\end{figure}

\subsection{Virial expansion of specific model}
In the simulation of \cite{Michieletto2016}, three kinds of marks are assumed ($1 \leq i \leq 3$): two of them ($i = 1$, $2$) with Ising-type interactions explained below and one of them ($i = 3$) with no mark-specific interactions.
For simplicity, we here neglect this third mark although its existence may shift the transition point slightly, as can the $O(\phi^6)$ terms.
The polymer model with two kinds of marks, where the interaction between a monomer with the mark $i$ and another monomer with the mark $j$ ($1 \leq i, j \leq 2$) is given as the truncated Lennard-Jones type~\cite{Michieletto2016}:
\begin{equation}
U_\mathrm{LJ}^{(ij)} (\bm{r}) = \frac{4 \epsilon^{(ij)}}{\mathcal{N}} \left[ \left( \frac{\sigma}{r} \right)^{12} - \left( \frac{\sigma}{r} \right)^6 - \left( \frac{\sigma}{r_\mathrm{c}^{(ij)}} \right)^{12} + \left( \frac{\sigma}{r_\mathrm{c}^{(ij)}} \right)^6 \right] \ \ (\mathrm{for} \ r \leq r_\mathrm{c}^{(ij)}),
\label{Eq:InteractionLJ}
\end{equation}
and $U_\mathrm{LJ}^{(ij)} (\bm{r}) = 0$ (for $r > r_\mathrm{c}^{(ij)}$).
Here, $\sigma$ corresponds to the interaction range, $\epsilon_{ij} = \epsilon \delta_{ij} + k_\mathrm{B} T (1 - \delta_{ij})$ represents the mark-specific interaction strength, $r_\mathrm{c}^{(ij)} = 1.8 \sigma \delta_{ij} + 2^{1/6} \sigma (1 - \delta_{ij})$ is the mark-specific interaction cutoff, and $\mathcal{N} = 1 + 4(1.8^{-12} - 1.8^{-6})$ is chosen so that $\min_{\bm{r}} U_\mathrm{LJ}^{(ii)} (\bm{r}) = -\epsilon$ [see Fig.~\ref{Fig:Virial}(a)].

Noticing the Ising symmetry of $U_\mathrm{LJ}^{(ij)} (\bm{r})$, i.e., $U_\mathrm{LJ}^{(11)} (\bm{r}) = U_\mathrm{LJ}^{(22)} (\bm{r})$, we identify $U_\mathrm{LJ}^{(ij)} (\bm{r})$ as $U (\bm{r}; i, j)$ in Sec.~\ref{Subsec:RelationVirial} and calculate the dimensionless coupling constants $\tilde{v}$, $\tilde{J}$, and $\tilde{w}'$ based on Eqs.~\eqref{Eq:Formula_v}, \eqref{Eq:Formula_J}, \eqref{Eq:Formula_w}, and \eqref{Eq:Formula_wprime}.
Since the Ising-type interaction strength can be quantified by a dimensionless parameter $\epsilon / k_\mathrm{B} T$, we plot $\tilde{v}$, $\tilde{J}$, and $\tilde{w}'$ as a function of $\epsilon / k_\mathrm{B} T$ in Fig.~\ref{Fig:Virial}(b).
The mark-specific interaction strength $\tilde{J}$ is an increasing function of $\epsilon / k_\mathrm{B} T$ as expected, while $\tilde{v}$ and $\tilde{w}'$ also depend on the value of $\epsilon / k_\mathrm{B} T$ since $\epsilon$ also affects the mark-nonspecific interactions.

By minimizing Eq.~\eqref{Eq:PseudoFreeEnergy_Virial}, we can obtain the equilibrium globular order parameter $\phi^*$ and magnetization $m^*$, as shown in Fig.~\ref{Fig:Virial}(c) as a function of $\epsilon / k_\mathrm{B} T$.
There is a first-order transition between the swollen-disordered state ($\phi^* = 0$, $m^* = 0$) and the compact-ordered state ($\phi^* > 0$, $|m^*| > 0$) at $\epsilon / k_\mathrm{B} T \simeq 0.85$, which is of the same order as the critical value ($\epsilon / k_\mathrm{B} T \simeq 0.9$) obtained in the simulation with $N = 2000$~\cite{Michieletto2016}.

Lastly, we stress that the necessary condition for a second-order coil-globule transition derived in the main text [see the discussions around Eq.~(6)] is applicable even when $m^2 \phi^4$ terms or $O (\phi^6)$ terms are present in the pseudo free energy as Eq.~\eqref{Eq:PseudoFreeEnergy_Virial}.
Indeed, $\tilde{v}$ is positive around the transition point [see Figs.~\ref{Fig:Virial}(b) and (c)], and thus the second-order coil-globule transition is prohibited, consistent with the numerical result.

\end{document}